\documentclass[12pt,draftclsnofoot,onecolumn]{IEEEtran}
\usepackage{cite}
\usepackage{amsmath}
\usepackage{amssymb}
\usepackage{graphicx}
\usepackage{epstopdf}
\usepackage{makecell}
\usepackage{color}
\usepackage{subcaption}

\begin{document}
\title{Backscatter Cooperation in NOMA Communications Systems}

\author{Weiyu~Chen,
        Haiyang~Ding, \IEEEmembership{Member,~IEEE,}
        Shilian~Wang, \IEEEmembership{Member,~IEEE,}\\
        Daniel~Benevides~da~Costa, \IEEEmembership{Senior~Member,~IEEE,}
        Fengkui~Gong,
        and~Pedro~Henrique~Juliano~Nardelli

\thanks{\scriptsize{W. Chen and S. Wang are with College of Electronic Science and Technology, National University of Defense Technology, Changsha, China (email: chenweiyu14@nudt.edu.cn, wangsl@nudt.edu.cn). This work was supported in part by the National Key R\&D Program of China under Grant 2018YFE0100500; by the National Natural Science Foundation of China under Grant 61871387, Grant 61861041, and Grant 61871471; by the Natural Science Basic Research Program of Shaanxi under Grant 2019JM-019; by Academy of Finland via: (a) ee-IoT n.319009, (b) FIREMAN consortium CHIST-ERA/n.326270, and (c) EnergyNet Research Fellowship n.321265/n.328869; and by the NUDT Research Fund under Grant ZK17-03-08.}}
\thanks{\scriptsize{H. Ding is with College of Information and Communication, National University of Defense Technology, Xi'an, China (email: dinghy2003@hotmail.com).}}
\thanks{\scriptsize{D. B. da Costa is with the Department of Computer Engineering, Federal University of Cear\'a, Sobral, CE, Brazil (email: danielbcosta@ieee.org).}}
\thanks{\scriptsize{F. Gong is with the State Key Laboratory of Integrated Service Networks, Xidian University, Xi'an, China (e-mail: fkgong@xidian.edu.cn).}}
\thanks{\scriptsize{P. H. J. Nardelli is with the School of Energy Systems, Lappeenranta University of Technology, Lappeenranta, Finland (e-mail: Pedro.Juliano.Nardelli@lut.fi).}}
\thanks{\scriptsize{Corresponding author: Shilian~Wang.}}
}

\maketitle

\IEEEpeerreviewmaketitle

\begin{abstract}
In this paper, a backscatter cooperation (BC) scheme is proposed for non-orthogonal multiple access (NOMA) downlink transmission. The key idea is to enable one user to split and then backscatter part of its received signals to improve the reception at another user. To evaluate the performance of the proposed BC-NOMA scheme, three benchmark schemes are introduced. They are the non-cooperation (NC)-NOMA scheme, the conventional relaying (CR)-NOMA scheme, and the incremental relaying (IR)-NOMA scheme. For all these schemes, the analytical expressions of the minimum total power to avoid information outage are derived, based on which their respective outage performance, expected rates, and diversity-multiplexing trade-off (DMT) are investigated. Analytical results show that the proposed BC-NOMA scheme strictly outperforms the NC-NOMA scheme in terms of all the three metrics. Furthermore, theoretical analyses are validated via Monte-Carlo simulations. It is shown that unlike the CR-NOMA scheme and the IR-NOMA scheme, the proposed BC-NOMA scheme can enhance the transmission reliability without impairing the transmission rate, which makes backscattering an appealing solution to cooperative NOMA downlinks.
\end{abstract}


\section{Introduction}
\IEEEPARstart{N}{on-orthogonal} multiple access (NOMA), whose key idea is to allow multiple users to use the same channels (i.e., the same time, frequency, and code resources) to access the network, has been recognized as a promising technique to achieve a higher spectrum efficiency in the fifth-generation (5G) network \cite{Dai18CommSur}. As one of many specific techniques of NOMA, power-domain NOMA, which utilizes superposition coding at the transmitter and successive interference cancellation (SIC) at the receiver, has received a great deal of attention due to its high compatibility with other techniques and low implementation complexity \cite{Saito13VTC,Islam17CommSur}. Specifically, in power-domain NOMA\footnote{This paper focuses on power-domain NOMA, which we refer to as NOMA for conciseness in the rest of the paper.}, the users with a worse channel condition are allocated with a higher power level. In this way, the users with a better channel condition can firstly decode and subtract the intended signals for the users with a worse channel condition from their observations, and then recover their own information. Benefiting from this mechanism, NOMA can achieve a 30\% system-level performance improvement over orthogonal multiple access (OMA) \cite{Benjebbour13ISPACS}. Extensive studies have been done for NOMA \cite{Zhang16CL,Sadia18ELEKTRO,Liu18TWC,Ali19TC}. Particularly, the achievable sum data rate and the outage probability of NOMA uplink transmission were investigated in \cite{Zhang16CL}, whereas the bit error rates under different channel fading types for NOMA downlink transmission were analyzed in \cite{Sadia18ELEKTRO}. Furthermore, the authors in \cite{Liu18TWC} proposed a joint transmission scheme to coordinate multiple base stations (BSs) to improve the coverage and the throughput of heterogeneous NOMA cellular networks. The effects of different user clustering models and different user ordering methods on the performance of large-scale NOMA networks were studied in \cite{Ali19TC}.

Cooperative communication is an effective approach to enhance the reliability of communication systems by providing diversity \cite{Laneman04TIT}, which has been introduced in multi-user NOMA downlinks and was shown to be able to achieve a diversity order of $K$ at all the $K$ users \cite{DingZhiguo15CommuLet}. Furthermore, for a NOMA downlink scenario where no direct link exists between the BS and the cell-edge user, the work in \cite{Kim15CL} introduced a dedicated decode-and-forward (DF) relay to facilitate the reception of the cell-edge user. The authors in \cite{Ding16WCL} further investigated the impacts of relay selection strategies on the performance of relay-assisted NOMA downlinks, where the direct links between the BS and all the users are assumed to be blocked. In addition, the combination of simultaneous wireless information and power transfer (SWIPT) and cooperative NOMA has been extensively investigated \cite{Liu16JSAComm,Yang17}, since it is desirable to compensate the consumed energy for cooperation at the helping nodes by harvesting energy from the downlink signals.

Although the aforementioned cooperative NOMA schemes are capable of enhancing the reliability, they require additional time slots for the relaying operation. To address this problem, the authors in \cite{Zhou18TWC} introduced the dynamic DF scheme into NOMA downlink transmission, where one codeword spans several blocks within a time slot. Only when the cell-center user has successfully recovered the codeword before the end of the time slot, it helps forward cell-edge user's information. Another solution is to utilize the on/off scheme proposed in \cite{Do18TC}, where additional time slots for relaying are activated only when the channel condition of the direct link from the BS to the cell-edge user is not good enough. However, these solutions still sacrifice part of the time resources of the cell-center user. To avoid this issue, one promising approach is to adopt the full-duplex (FD) technology \cite{Zhang15CommMag}. In this regard, the authors in \cite{Zhong16CL} introduced a dedicated FD relay to fulfill the information transmission between the BS and the cell-edge user, whereas the work in \cite{Zhang17TVT} investigated a scenario where the cell-center user is a FD device and helps enhance the reception at the cell-edge user. However, FD relaying introduces non-negligible residual loop self-interference, which may impair the reception at the helping nodes.

On the other hand, ambient backscatter communication (AmBC) is emerging as a potential technique to improve both spectrum efficiency and energy efficiency for green Internet-of-Things (IoT) \cite{Liu13ACM,Huynh18CommuSur,Zheng19Access}. Specifically, in AmBC, the transmitter varies its load impedance to change the amplitude and/or phase of the backscattered signals to transmit information. Very recently, the backscatter technique was utilized to produce constructive multi-path signals to enhance the reliability of communication systems \cite{Gong18TC,Xu19IoT}. Particularly, the authors in \cite{Gong18TC} considered a network with multiple backscatter transceiver pairs and a power beacon station, in which idle backscatter transmitters backscatter the signals from the backscatter transmitter who occupies the time slot to improve the reception at the receiver. Similarly, an active transmission can also be assisted by idle transmitters with the help of the backscatter technique \cite{Xu19IoT}. Compared with FD cooperation, backscatter cooperation (BC) does not introduce self-interference and can provide diversity without sacrificing additional time resources. Meanwhile, BC does not require a local oscillator to generate carrier signals, which means that its power consumption is much lower than that of FD cooperation. In view of these potential benefits of backscatter transmission, in this paper, we incorporate BC into NOMA downlink systems to enhance the reception in a spectrally-efficient manner. The main contributions can be summarized as follows.

1) For NOMA downlink transmission, we propose a BC-NOMA scheme, in which the user with a better instantaneous channel condition splits its received signals into two parts. One part is used for information decoding, whereas the other part is backscattered to improve the reception at the user with a worse instantaneous channel condition.

2) For comparison purposes, three benchmark schemes are introduced, including the non-cooperation (NC)-NOMA scheme, the conventional relaying (CR)-NOMA scheme, and the incremental relaying (IR)-NOMA scheme. The close-form expressions of the minimum total power to avoid information outage are derived for all the four schemes, which show that the three cooperative schemes (i.e., CR, IR, and BC) indeed help reduce the minimum total power compared with the NC-NOMA scheme.

3) The outage performance, the expected rates, and the diversity-multiplexing trade-off (DMT) of the four schemes are analyzed under Rayleigh fading channels. Theoretical results show that the proposed BC-NOMA scheme strictly outperforms the NC-NOMA scheme in terms of all the three metrics, which means that it can enhance the reliability without impairing the effectiveness.

The rest of the paper is organized as follows. Section II illustrates the system model and different schemes. The theoretical analyses are presented in Section III. Section IV provides representative numerical results. Finally, Section V concludes the paper.
\section{System Model and Cooperative Schemes}
We consider a typical two-user NOMA downlink transmission, in which a BS transmits the information of two users (denoted by user A and user B, respectively) with different transmit power, $P_{\textrm{A}}$ and $P_{\textrm{B}}$, respectively. The system works in the delay-constrained transmission mode \cite{Zhong14TC} and the target data rates at user A and user B are $R_{\textrm{A}}$ and $R_{\textrm{B}}$, respectively. Quasi-static channels are considered, i.e., the channel coefficients $h_{\textrm{A}}$, $h_{\textrm{B}}$, and $g$ pertaining to the BS-A, BS-B, and A-B links remain unchanged within each transmission block (a.k.a. fading block), but may vary for different blocks. We assume that the BS maintains global channel state information (CSI), and denote $\lambda_{\textrm{A}}$, $\lambda_{\textrm{B}}$, and $\lambda_{\textrm{g}}$ as the means of $|h_{\textrm{A}}|^2$, $|h_{\textrm{B}}|^2$, and $|g|^2$, respectively.

Let $n_{\textrm{A}}$ and $n_{\textrm{B}}$ represent the zero-mean additive white Gaussian noise (AWGN) at A and B with variances $\sigma_{\textrm{A}}^2$ and $\sigma_{\textrm{B}}^2$, respectively. To proceed, we denote the user with a better instantaneous channel condition as user 1 and that with a worse instantaneous channel condition as user 2, respectively. Specifically, for each fading block, if $|h_{\textrm{A}}|^2/\sigma_{\textrm{A}}^2\ge |h_{\textrm{B}}|^2/\sigma_{\textrm{B}}^2$, we define $P_1\triangleq P_{\textrm{A}}$, $P_2\triangleq P_{\textrm{B}}$, $\sigma_{1}^2\triangleq \sigma_{\textrm{A}}^2$, $\sigma_{2}^2\triangleq \sigma_{\textrm{B}}^2$, $n_1\triangleq n_{\textrm{A}}$, $n_2\triangleq n_{\textrm{B}}$, $R_1\triangleq R_{\textrm{A}}$, $R_2\triangleq R_{\textrm{B}}$, $h_1\triangleq h_{\textrm{A}}$, and $h_2\triangleq h_{\textrm{B}}$. By its turn, if $|h_{\textrm{A}}|^2/\sigma_{\textrm{A}}^2< |h_{\textrm{B}}|^2/\sigma_{\textrm{B}}^2$, we define $P_1\triangleq P_{\textrm{B}}$, $P_2\triangleq P_{\textrm{A}}$, $\sigma_{1}^2\triangleq \sigma_{\textrm{B}}^2$, $\sigma_{2}^2\triangleq \sigma_{\textrm{A}}^2$, $n_1\triangleq n_{\textrm{B}}$, $n_2\triangleq n_{\textrm{A}}$, $R_1\triangleq R_{\textrm{B}}$, $R_2\triangleq R_{\textrm{A}}$, $h_1\triangleq h_{\textrm{B}}$, and $h_2\triangleq h_{\textrm{A}}$. Based on these definitions, we can arrive at a unified description of the schemes discussed in the following.

\subsection{NC-NOMA Scheme}
In a basic two-user NOMA downlink transmission, the BS superposes and broadcasts the information of the two users over the same spectrum. The received signals at user 1 and user 2 can be written, respectively, as \cite{Saito13VTC}
\begin{align}
y_{1}=\left(\sqrt{P_{1}} x_{1}+\sqrt{P_{2}} x_{2}\right) h_{1}+n_{1},
\label{E1_y1}
\end{align}
\begin{align}
y_{2}=\left(\sqrt{P_{1}} x_{1}+\sqrt{P_{2}} x_{2}\right) h_{2}+n_{2},
\label{E2_y2}
\end{align}
where $x_1$ and $x_2$ denote the normalized intended signals for user 1 and user 2 (i.e., $E\{|x_1 |^2 \}=E\{|x_2 |^2 \}=1$), respectively. Hereafter, when either user 1 cannot recover $x_1$ or user 2 cannot recover $x_2$, we say that an information outage happens, and the corresponding probability is called system outage probability (SOP).

According to the principles of NOMA, user 2 is allocated with a higher power level (i.e., $P_2\ge P_1$). User 1 first decodes $x_2$ by treating $x_1$ as noise, and then subtracts $x_2$ from its received signals to decode $x_1$, whereas user 2 only needs to decode $x_2$ by treating $x_1$ as noise. The received signal-to-interference-plus-noise ratio (SINR) at user 1 and user 2 to decode $x_2$ can be given, respectively, as
\begin{align}
\gamma_{12}=\frac{P_{2}\left|h_{1}\right|^{2}}{P_{1}\left|h_{1}\right|^{2}+\sigma_{1}^{2}},
\label{E3_gamma12}
\end{align}
\begin{align}
\gamma_{22}=\frac{P_{2}\left|h_{2}\right|^{2}}{P_{1}\left|h_{2}\right|^{2}+\sigma_{2}^{2}}.
\label{E4_gamma22}
\end{align}
Correspondingly, the inequalities $\gamma_{12}\ge \overline{\gamma}_{2}$ and $\gamma_{22}\ge \overline{\gamma}_{2}$ are the conditions for user 1 and user 2 to successfully decode $x_2$, respectively, where $\overline{\gamma}_{2}\triangleq2^{R_2}-1$. After successfully decoding $x_2$, the signal-to-noise ratio (SNR) at user 1 to decode $x_1$ can be written as 
\begin{align}
\gamma_{11}=\frac{P_{1}\left|h_{1}\right|^{2}}{\sigma_{1}^{2}}.
\label{E5_gamma11}
\end{align}
When $\gamma_{11}\ge \overline{\gamma}_{1}$, user 1 decodes $x_1$ successfully, where $\overline{\gamma}_{1}\triangleq2^{R_1}-1$.
\subsection{CR-NOMA Scheme}
There are two phases in the CR-NOMA scheme, i.e., the direct transmission phase and the cooperative transmission phase \cite{DingZhiguo15CommuLet}. In the direct transmission phase, the received signals at user 1 and user 2 can still be given by \eqref{E1_y1} and \eqref{E2_y2}, respectively. Also, for user 1, the SINR to decode $x_2$ and the SNR to decode $x_1$ can be still written as in \eqref{E3_gamma12} and \eqref{E5_gamma11}, respectively. The difference between the NC-NOMA scheme and the CR-NOMA scheme lies in that user 1 forwards $x_2$ to user 2 in the cooperative phase for the CR-NOMA scheme. The corresponding received signals at user 2 can be represented as
\begin{align}
y_{2, \textrm{CT}}=\sqrt{P_{\textrm{h}}} x_{2} g+n_{2},
\label{E6_y2CT}
\end{align}
where $P_{\textrm{h}}$ denotes the transmit power at user 1. In this paper, a peak total power constraint is considered, which can be written as $P_1+P_2+P_{\textrm{h}}\le P_{\textrm{p}}$, where $P_{\textrm{p}}$ denotes the maximum allowed total power. Next, by applying the maximal-ratio combining (MRC) technique \cite{Goldsmith05Wireless}, user 2 combines its received signals of the two phases to decode $x_2$, and the corresponding SINR can be written as
\begin{align}
\gamma_{22, \textrm{MRC}}=\frac{P_{2}\left|h_{2}\right|^{2}}{P_{1}\left|h_{2}\right|^{2}+\sigma_{2}^{2}}+\frac{P_{\textrm{h}}|g|^{2}}{\sigma_{2}^{2}}.
\label{E7_gamma22MRC}
\end{align}
As before, the conditions for user 1 to decode $x_2$ and $x_1$ successfully are $\gamma_{12}\ge \overline{\gamma}_{2}$ and $\gamma_{11}\ge \overline{\gamma}_{1}$, whereas the condition for user 2 to decode $x_2$ successfully becomes $\gamma_{22, \textrm{MRC}}\ge \overline{\gamma}_{2}$.
\subsection{IR-NOMA Scheme}
Unlike the CR-NOMA scheme, the cooperative transmission phase is not indispensable for the IR-NOMA scheme \cite{Do18TC}. Specifically, to maximize the spectrum efficiency, if information outage can be avoided in the direct transmission phase by properly allocating the transmit power at the BS under the peak total power constraint, or if information outage cannot be avoided even if the cooperative transmission phase is activated, the cooperative transmission phase will not be activated for the IR-NOMA scheme.

For the fading blocks when the cooperative transmission is activated, the received SINR/SNR expressions at both users are the same as the counterparts in the CR-NOMA scheme. Otherwise, they are the same as those in the NC-NOMA scheme.

Note that for both the CR-NOMA scheme and the IR-NOMA scheme, cooperation is conducted at the cost of introducing extra time slots, which may impair the data rate. This motivates us to propose the BC-NOMA scheme as follows.
\subsection{BC-NOMA Scheme}
\begin{figure}[htp]
	\centering
	\includegraphics[scale=1.16]{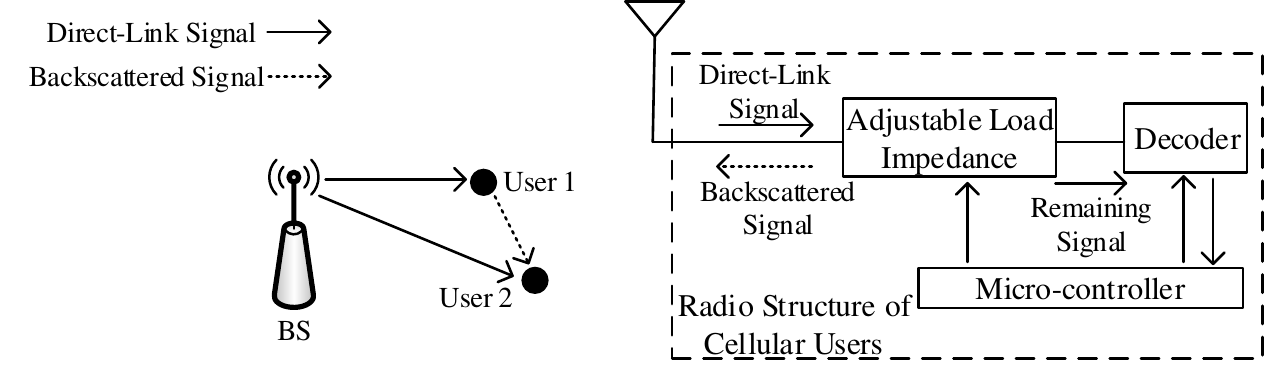}
	\caption{BC-NOMA scheme.}
	\label{Fig1BCNOMAScheme}
\end{figure}
Fig. \ref{Fig1BCNOMAScheme} illustrates the proposed BC-NOMA scheme. Specifically, by adjusting the load impedance, user 1 splits part of its received signals for information decoding, and backscatters the remaining part to user 2 to improve the SINR at user 2. The received signals at user 1 can still be given by \eqref{E1_y1} for the BC-NOMA scheme, whereas the received signals at user 2 can be written as\footnote{As in \cite{Gong18TC,Xu19IoT}, we assume that the delay of the backscattered signals from user 1 with respect to the direct-link signals from the BS is negligible. Note that this assumption is reasonable in many scenarios. For example, in a small cell, the two users are typically close to each other and thus the delay is negligible. Also, for a low-rate transmission, the symbol period is typically much longer than the delay such that the delay becomes negligible.}\footnote{We assume that the backscattered noise is negligible due to the passive nature of backscatter circuits, as in \cite{Gong18TC,Xu19IoT}.}
\begin{align}
y_{2, \textrm{BT}}=\left(\sqrt{P_{1}} x_{1}+\sqrt{P_{2}} x_{2}\right) h_{2}+\sqrt{\beta_{1}}\left(\sqrt{P_{1}} x_{1}+\sqrt{P_{2}} x_{2}\right) h_{1} g+n_{2},
\label{E8_y2BT}
\end{align}
where $\beta_{1}\in[0,1)$ denotes the percentage of the received power backscattered by user 1. To simplify the presentation, we incorporate the backscatter efficiency $\eta$ ($0<\eta\le1$) due to the imperfectness of circuit implementation into the channel coefficient $g$ in \eqref{E8_y2BT}, where $g\triangleq \sqrt{\eta} g^{\prime}$ and $g^{\prime}$ denotes the original channel coefficient. Next, for user 2, the SINR to decode $x_2$ can be given by
\begin{align}
\gamma_{22, \textrm{BT}}=\frac{P_{2}\left|h_{2}+\sqrt{\beta_{1}} h_{1} g\right|^{2}}{P_{1}\left|h_{2}+\sqrt{\beta_{1}} h_{1} g\right|^{2}+\sigma_{2}^{2}}
\underset{\text{A}}{>}
\frac{P_{2}\left(\left|h_{2}\right|^{2}+\beta_{1}\left|h_{1}\right|^{2}|g|^{2}\right)}{P_{1}\left(\left|h_{2}\right|^{2}+\beta_{1}\left|h_{1}\right|^{2}|g|^{2}\right)+\sigma_{2}^{2}}
\triangleq
\gamma_{22, \textrm{BT}}^{\prime}.
\label{E9_gamma22BT}
\end{align}
As in \cite{Gong18TC,Xu19IoT}, herein the helping node (user 1) adjusts its complex reflection coefficient to generate constructive multi-path signals at the helped node (user 2), and the best strategy at user 1 is to adjust its reflection coefficient to make the received backscattered signals at user 2 have the same phase as that of the received direct-link signals at user 2. As thus, step A in \eqref{E9_gamma22BT} holds. To simplify the problem, hereafter we use $\gamma_{22, \textrm{BT}}^{\prime}$ as the received SINR at user 2 and denote it as $\gamma_{22, \textrm{BT}}$, based on which the derived performance of the BC-NOMA scheme can be regarded as its strict lower bound.

On the other hand, for user 1, the SINR to decode $x_2$ and the SNR to decode $x_1$ can be written, respectively, as
\begin{align}
\gamma_{12, \textrm{BT}}=\frac{P_{2}\left|h_{1}\right|^{2}\left(1-\beta_{1}\right)}{P_{1}\left|h_{1}\right|^{2}\left(1-\beta_{1}\right)+\sigma_{1}^{2}},
\label{E10_gamma12BT}
\end{align}
\begin{align}
\gamma_{11, \textrm{BT}}=\frac{P_{1}\left|h_{1}\right|^{2}\left(1-\beta_{1}\right)}{\sigma_{1}^{2}}.
\label{E11_gamma11BT}
\end{align}

\section{Performance Analysis and Comparisons}
In this section, for each of the four schemes, the close-form expressions of the minimum required total power to avoid information outage for a given fading block are derived, based on which the SOP and the expected rate are investigated to evaluate the reliability and the effectiveness, respectively. Finally, the DMT performance is analyzed to further reveal the trade-off between the reliability and the effectiveness in the high SNR region.
\subsection{Minimum Required Total Power}
\subsubsection{NC-NOMA Scheme}
The minimum required total power for the NC-NOMA scheme can be derived by solving the following problem.
\begin{align}
\text{P1}:\underset{P_1,P_2}{\textrm{min}}&P_1+P_2,\nonumber\\
\text{s. t. }& \gamma_{11} \geq \overline{\gamma}_{1}, \gamma_{12} \geq \overline{\gamma}_{2}, \gamma_{22} \geq \overline{\gamma}_{2}.
\label{E12_P1}
\end{align}

\textbf{\emph{Proposition 1}}: For the NC-NOMA scheme, the minimum required total power to avoid information outage can be written as $P_{\textrm{min,NC}} =\frac{\sigma_{1}^{2}(\overline{\gamma}_{1}+\overline{\gamma}_{1} \overline{\gamma}_{2})}{\left|h_{1}\right|^{2}}+\frac{\sigma_{2}^{2} \overline{\gamma}_{2}}{\left|h_{2}\right|^{2}}$.

\emph{Proof}: By expanding and then combining the constraints of P1, one can readily arrive at Proposition 1. The proof is trivial and thus is omitted here.\hfill$\blacksquare$
\subsubsection{CR-NOMA Scheme}
The minimum required total power for the CR-NOMA scheme can be derived by solving the following problem.
\begin{align}
\text{P2}:\underset{P_1,P_2,P_{\textrm{h}}}{\textrm{min}}&P_1+P_2+P_{\textrm{h}},\nonumber\\
\text{s. t. }& \gamma_{11} \geq \overline{\gamma}_{1}, \gamma_{12} \geq \overline{\gamma}_{2}, \gamma_{22, \textrm{MRC}} \geq \overline{\gamma}_{2}.
\label{E13_P2}
\end{align}

\textbf{\emph{Proposition 2}}: For the CR-NOMA scheme, the minimum required total power to avoid information outage can be written as
\begin{align}
P_{\textrm{min,CR}}=
\begin{cases}
\frac{\sigma_{1}^{2}\left(\overline{\gamma}_{1}+\overline{\gamma}_{1} \overline{\gamma}_{2}\right)}{\left|h_{1}\right|^{2}}+\frac{\sigma_{2}^{2} \overline{\gamma}_{2}}{\left|h_{2}\right|^{2}},&|g|^{2} \leq\left(\frac{1}{\left|h_{2}\right|^{2}}+\frac{\sigma_{1}^{2} \overline{\gamma}_{1}}{\sigma_{2}^{2}\left|h_{1}\right|^{2}}\right)^{-1},\\
\frac{\sigma_{1}^{2}\left(\overline{\gamma}_{1}+\overline{\gamma}_{1} \overline{\gamma}_{2}+\overline{\gamma}_{2}\right)}{\left|h_{1}\right|^{2}}+\frac{\overline{\gamma}_{2}\left(\frac{\sigma_{2}^{2}}{\left|h_{2}\right|^{2}}-\frac{\sigma_{1}^{2}}{\left|h_{1}\right|^{2}}\right)}{|g|^{2}\left(\frac{\sigma_{1}^{2} \overline{\gamma}_{1}}{\sigma_{2}^{2}\left|h_{1}\right|^{2}}+\frac{1}{\left|h_{2}\right|^{2}}\right)},&|g|^{2} >\left(\frac{1}{\left|h_{2}\right|^{2}}+\frac{\sigma_{1}^{2} \overline{\gamma}_{1}}{\sigma_{2}^{2}\left|h_{1}\right|^{2}}\right)^{-1}.
\end{cases}
\label{E14_PminTR}
\end{align}

\emph{Proof}: Please refer to Appendix A-1.\hfill$\blacksquare$

\emph{Remark 1}: Note that when $|g|^{2} \leq\left(\frac{1}{\left|h_{2}\right|^{2}}+\frac{\sigma_{1}^{2} \overline{\gamma}_{1}}{\sigma_{2}^{2}\left|h_{1}\right|^{2}}\right)^{-1}$, we have $P_{\textrm{min,CR}}=P_{\textrm{min,NC}}$. On the other hand, when $|g|^{2} >\left(\frac{1}{\left|h_{2}\right|^{2}}+\frac{\sigma_{1}^{2} \overline{\gamma}_{1}}{\sigma_{2}^{2}\left|h_{1}\right|^{2}}\right)^{-1}$, it is easy to validate that $P_{\textrm{min,CR}}<P_{\textrm{min,NC}}$, which means that the introduction of the cooperative transmission phase does help reduce the minimum required total power, and this happens when the channel power gain of the cooperative channel (i.e., $|g|^2$) is high enough.

\subsubsection{IR-NOMA Scheme}
The minimum required total power for the IR-NOMA scheme is the same as that for the CR-NOMA scheme (i.e., $P_{\textrm{min,IR}}=P_{\textrm{min,CR}}$). This is because both schemes are able to activate the cooperative transmission phase to enhance the SINR at user 2.

Remember that if information outage can be avoided in the direct transmission phase by properly allocating the transmit power at the BS under the peak total power constraint, the cooperative transmission phase will not be activated for the IR-NOMA scheme. In other words, when $P_{\textrm{min,NC}}\le P_{\textrm{p}}$, the cooperative transmission phase will not be activated even if it helps reduce the required total power to avoid information outage, and thus in this case the required total power is $P_{\textrm{min,NC}}$, instead of $P_{\textrm{min,IR}}$.

\subsubsection{BC-NOMA Scheme}
The minimum required total power for the BC-NOMA scheme can be derived by solving the following problem.
\begin{align}
\text{P3}:\underset{P_1,P_2,\beta_{1}}{\textrm{min}}&P_1+P_2,\nonumber\\
\text{s. t. }& \gamma_{11, \textrm{BT}} \geq \overline{\gamma}_{1}, \gamma_{12, \textrm{BT}} \geq \overline{\gamma}_{2}, \gamma_{22, \textrm{BT}} \geq \overline{\gamma}_{2}.
\label{E15_P3}
\end{align}

\textbf{\emph{Proposition 3}}: For the BC-NOMA scheme, when $|g|^{2} \leq\left(\frac{\left|h_{2}\right|^{2}}{\left|h_{1}\right|^{2}}\right)^{2} \frac{\sigma_{1}^{2}(\overline{\gamma}_{1}+\overline{\gamma}_{1}\overline{\gamma}_{2})}{\sigma_{2}^{2} \overline{\gamma}_{2}}$, the minimum required total power to avoid information outage is $P_{\textrm{min,BC}}=\frac{\sigma_{1}^{2}(\overline{\gamma}_{1}+\overline{\gamma}_{1}\overline{\gamma}_{2})}{\left|h_{1}\right|^{2}}+\frac{\sigma_{2}^{2} \overline{\gamma}_{2}}{\left|h_{2}\right|^{2}}$. When $|g|^{2} \geq \frac{\sigma_{2}^{2}(\overline{\gamma}_{1}+\overline{\gamma}_{1}\overline{\gamma}_{2})}{\sigma_{1}^{2} \overline{\gamma}_{2}}$, we have $P_{\textrm{min,BC}}=\frac{(\overline{\gamma}_{1}+\overline{\gamma}_{1}\overline{\gamma}_{2}+\overline{\gamma}_{2})\left(\sigma_{2}^{2}+\sigma_{1}^{2}|g|^{2}\right)}{\left|h_{1}\right|^{2}|g|^{2}+\left|h_{2}\right|^{2}}$. When $\left(\frac{\left|h_{2}\right|^{2}}{\left|h_{1}\right|^{2}}\right)^{2} \frac{\sigma_{1}^{2}(\overline{\gamma}_{1}+\overline{\gamma}_{1}\overline{\gamma}_{2})}{\sigma_{2}^{2} \overline{\gamma}_{2}}<|g|^{2} < \frac{\sigma_{2}^{2}(\overline{\gamma}_{1}+\overline{\gamma}_{1}\overline{\gamma}_{2})}{\sigma_{1}^{2} \overline{\gamma}_{2}}$, it follows that
\begin{align}
P_{\textrm{min,BC}}=\frac{\sigma_{1}^{2}\left(\overline{\gamma}_{1}+\overline{\gamma}_{1} \overline{\gamma}_{2}\right)\left(2 \sqrt{\frac{\sigma_{2}^{2} \overline{\gamma}_{2}|g|^{2}}{\sigma_{1}^{2}\left(\overline{\gamma}_{1}+\overline{\gamma}_{1} \overline{\gamma}_{2}\right)}}+|g|^{2}\right)+\sigma_{2}^{2} \overline{\gamma}_{2}}{\left|h_{1}\right|^{2}|g|^{2}+\left|h_{2}\right|^{2}}.
\label{E16_PminBC3}
\end{align}

\emph{Proof}: Please refer to Appendix A-2.\hfill$\blacksquare$

\emph{Remark 2}: It follows from Propositions 1 and 3 that when $|g|^{2} \leq\left(\frac{\left|h_{2}\right|^{2}}{\left|h_{1}\right|^{2}}\right)^{2} \frac{\sigma_{1}^{2}(\overline{\gamma}_{1}+\overline{\gamma}_{1}\overline{\gamma}_{2})}{\sigma_{2}^{2} \overline{\gamma}_{2}}$, we have $P_{\textrm{min,BC}}=P_{\textrm{min,NC}}$. On the other hand, when $|g|^{2} >\left(\frac{\left|h_{2}\right|^{2}}{\left|h_{1}\right|^{2}}\right)^{2} \frac{\sigma_{1}^{2}(\overline{\gamma}_{1}+\overline{\gamma}_{1}\overline{\gamma}_{2})}{\sigma_{2}^{2} \overline{\gamma}_{2}}$, one can validate that $P_{\textrm{min,BC}}<P_{\textrm{min,NC}}$, which means that the introduction of the BC does help reduce the minimum required total power, and this happens when the channel power gain of the cooperative channel (i.e., $|g|^2$) is high enough.

\subsection{System Outage Probability}
As stated in Section II, the SOP is defined as the probability that either user 1 cannot recover $x_1$ or user 2 cannot recover $x_2$ with the total power consumption less than or equal to $P_{\textrm{p}}$. Note that this happens when $P_{\textrm{min}}>P_{\textrm{p}}$, where $P_{\textrm{min}}$ denotes $P_{\textrm{min,NC}}$, $P_{\textrm{min,CR}}$, $P_{\textrm{min,IR}}$, and $P_{\textrm{min,BC}}$ for the NC-NOMA scheme, the CR-NOMA scheme, the IR-NOMA scheme, and the BC-NOMA scheme, respectively. Therefore, the SOP can be written as\footnote{Remember that if $|h_{\textrm{A}}|^2/\sigma_{\textrm{A}}^2\ge |h_{\textrm{B}}|^2/\sigma_{\textrm{B}}^2$, we define $P_1\triangleq P_{\textrm{A}}$, $P_2\triangleq P_{\textrm{B}}$, $\sigma_{1}^2\triangleq \sigma_{\textrm{A}}^2$, $\sigma_{2}^2\triangleq \sigma_{\textrm{B}}^2$, $n_1\triangleq n_{\textrm{A}}$, $n_2\triangleq n_{\textrm{B}}$, $R_1\triangleq R_{\textrm{A}}$, $R_2\triangleq R_{\textrm{B}}$, $h_1\triangleq h_{\textrm{A}}$, and $h_2\triangleq h_{\textrm{B}}$. By its turn, if $|h_{\textrm{A}}|^2/\sigma_{\textrm{A}}^2< |h_{\textrm{B}}|^2/\sigma_{\textrm{B}}^2$, we define $P_1\triangleq P_{\textrm{B}}$, $P_2\triangleq P_{\textrm{A}}$, $\sigma_{1}^2\triangleq \sigma_{\textrm{B}}^2$, $\sigma_{2}^2\triangleq \sigma_{\textrm{A}}^2$, $n_1\triangleq n_{\textrm{B}}$, $n_2\triangleq n_{\textrm{A}}$, $R_1\triangleq R_{\textrm{B}}$, $R_2\triangleq R_{\textrm{A}}$, $h_1\triangleq h_{\textrm{B}}$, and $h_2\triangleq h_{\textrm{A}}$. Therefore, in \eqref{E17_Pout}, the expression of $P_{\textrm{min}}$ in the first term is essentially different from that in the second term.}
\begin{align}
P_{\textrm{out}}=\operatorname{Pr}\left(P_{\textrm{min}}>P_{\textrm{p}}, \frac{\left|h_{\textrm{A}}\right|^{2}}{\sigma_{\textrm{A}}^{2}} \geq \frac{\left|h_{\textrm{B}}\right|^{2}}{\sigma_{\textrm{B}}^{2}}\right)+\operatorname{Pr}\left(P_{\textrm{min}}>P_{\textrm{p}}, \frac{\left|h_{\textrm{A}}\right|^{2}}{\sigma_{\textrm{A}}^{2}}<\frac{\left|h_{\textrm{B}}\right|^{2}}{\sigma_{\textrm{B}}^{2}}\right).
\label{E17_Pout}
\end{align}
Since we have $P_{\textrm{min,IR}}=P_{\textrm{min,CR}}$, the SOP of the IR-NOMA scheme is the same as that of the CR-NOMA scheme. However, the expected rates and the DMT performance of the two schemes are not the same, which will be investigated in Section III-C and Section III-D, respectively. On the other hand, note that the derived expressions of the minimum required total power as well as \eqref{E17_Pout} hold regardless of the channel fading types. Therefore, we can arrive at the following conclusion.

\emph{Corollary 1}: The SOPs of the three cooperative schemes (i.e., CR, IR, and BC) are lower than or at most the same as that of the NC-NOMA scheme regardless of the channel fading types.

\emph{Proof}: It follows from Remarks 1 and 2 that the minimum required total power of the three cooperative schemes is no larger than that of the NC-NOMA scheme. Combining this observation with the definition of SOP in \eqref{E17_Pout}, we can draw the conclusion.\hfill$\blacksquare$

The detailed derivation of the SOP under Rayleigh fading channels is presented in Appendix B for the four schemes, which leads to involved expressions and does not give us any insight. To address this, based on the derived expressions, we will further examine the DMT performance of the four schemes in Section III-D, whereas the derived expressions will be validated and the SOPs of the four schemes will be compared in Section IV via numerical experiments.
\subsection{Expected Sum Rate}
The SOP characterizes the reliability of different schemes. Herein, we further use the expected sum rate (ESR) to evaluate the effectiveness, which is defined as the expectation of the sum data rate that the two users achieve. By taking the rate loss due to both the extra time slots and the information outage into account, the ESR can be written as\cite{Makki12TWC}
\begin{align}
R_{\textrm{ES}}=\frac{\left(1-P_{\textrm{out }}\right)\left(R_{\textrm{A}}+R_{\textrm{B}}\right) N_{\textrm{DT}}}{N_{\textrm{DT}}+N_{\textrm{CT}}}=\frac{\left(1-P_{\textrm{out}}\right)\left(R_{\textrm{A}}+R_{\textrm{B}}\right)}{1+P_{\textrm{CT}}},
\label{E18_RES}
\end{align}
where $N_{\textrm{DT}}$ and $N_{\textrm{CT}}$ represent the number of the direct transmission time slots and that of the cooperative transmission time slots in a long-term operation, respectively, whereas $P_{\textrm{CT}}$ denotes the probability that the cooperative transmission time slot is activated. Note that $P_{\textrm{CT}}=0$ for both the NC-NOMA scheme and the BC-NOMA scheme. For the CR-NOMA scheme, we have $P_{\textrm{CT}}=1$, whereas for the IR-NOMA scheme, $P_{\textrm{CT}}$ can be written as $\operatorname{Pr}\left(P_{\textrm{min,NC}}>P_{\textrm{p}}, P_{\textrm{min,IR}} \leq P_{\textrm{p}}\right)$.
This is because for the IR-NOMA scheme, the extra time slot is activated only if information outage cannot be avoided without the cooperative transmission phase but can be avoided with the help of the cooperative transmission phase. Note that \eqref{E18_RES} holds regardless of the channel fading types. Therefore, we can arrive at the following conclusion.

\emph{Corollary 2}: The ESR of the BC-NOMA scheme is higher than or at least the same as the counterpart of the NC-NOMA scheme regardless of the channel fading types.

\emph{Proof}: It follows from Corollary 1 that the SOP of the BC-NOMA scheme is no larger than that of the NC-NOMA scheme. On the other hand, note that $P_{\textrm{CT}}=0$ for both the BC-NOMA scheme and the NC-NOMA scheme. Combining these two observations with the definition of the ESR in \eqref{E18_RES}, we can arrive at Corollary 2.\hfill$\blacksquare$

As mentioned above, the expressions of $P_{\textrm{out}}$ of the four schemes under Rayleigh fading channels are presented in Appendix B. Therefore, to determine the ESRs of the four schemes under Rayleigh fading channels, we only need to derive the expression of $\operatorname{Pr}\left(P_{\textrm{min,NC}}>P_{\textrm{p}}, P_{\textrm{min,IR}} \leq P_{\textrm{p}}\right)$. For such, we define $\overline{\gamma}_\textrm{A}\triangleq2^{R_\textrm{A}}-1$ and $\overline{\gamma}_\textrm{B}\triangleq2^{R_\textrm{B}}-1$. By making use of Propositions 1 and 2, we have
	$\operatorname{Pr}\left(P_{\textrm{min,NC}}>P_{\textrm{p}}, P_{\textrm{min,IR}} \leq P_{\textrm{p}}\right)=
	\int_{\frac{\sigma_\textrm{A}^{2}\left( \overline{\gamma}_\textrm{A}+\overline{\gamma}_\textrm{A}\overline{\gamma}_\textrm{B}+\overline{\gamma}_\textrm{B}\right) }{P_{\textrm{p}}}}^{\infty} \int_{0}^{\frac{\sigma_\textrm{B}^{2} \overline{\gamma}_\textrm{B}}{\left(P_{\textrm{p}}-\frac{\sigma_\textrm{A}^{2}(\overline{\gamma}_\textrm{A}+\overline{\gamma}_\textrm{A}\overline{\gamma}_\textrm{B})}{x}\right)}}
	\frac{e^{-\frac{y}{\lambda_\textrm{B}}-\frac{\zeta_{1}}{\lambda_\textrm{g}}}}{\lambda_\textrm{B}} d y \frac{e^{-\frac{x}{\lambda_\textrm{A}}}}{\lambda_\textrm{A}} d x%
	+
	\int_{\frac{\sigma_\textrm{B}^{2}\left( \overline{\gamma}_\textrm{B}+\overline{\gamma}_\textrm{B}\overline{\gamma}_\textrm{A}+\overline{\gamma}_\textrm{A}\right) }{P_{\textrm{p}}}}^{\infty} \int_{0}^{\frac{\sigma_\textrm{A}^{2} \overline{\gamma}_\textrm{A}}{\left(P_{\textrm{p}}-\frac{\sigma_\textrm{B}^{2}(\overline{\gamma}_\textrm{B}+\overline{\gamma}_\textrm{B}\overline{\gamma}_\textrm{A})}{x}\right)}}
	\frac{e^{-\frac{y}{\lambda_\textrm{A}}-\frac{\zeta_{2}}{\lambda_\textrm{g}}}}{\lambda_\textrm{A}} d y \frac{e^{-\frac{x}{\lambda_\textrm{B}}}}{\lambda_\textrm{B}} d x$, where $\zeta_{1}\triangleq \frac{\sigma_\textrm{B}^{2} \overline{\gamma}_\textrm{B}\left(\frac{\sigma_\textrm{B}^{2}}{y}-\frac{\sigma_\textrm{A}^{2}}{x}\right)}{\left(P_{\textrm{p}}-\frac{\sigma_\textrm{A}^{2}\left(\overline{\gamma}_\textrm{A}+\overline{\gamma}_\textrm{A}\overline{\gamma}_\textrm{B}+\overline{\gamma}_\textrm{B}\right)}{x}\right)\left(\frac{\sigma_\textrm{B}^{2}}{y}+\frac{\sigma_\textrm{A}^{2} \overline{\gamma}_\textrm{A}}{x}\right)}$ and
	$\zeta_{2}\triangleq \frac{\sigma_\textrm{A}^{2} \overline{\gamma}_\textrm{A}\left(\frac{\sigma_\textrm{A}^{2}}{y}-\frac{\sigma_\textrm{B}^{2}}{x}\right)}{\left(P_{\textrm{p}}-\frac{\sigma_\textrm{B}^{2}\left(\overline{\gamma}_\textrm{B}+\overline{\gamma}_\textrm{B}\overline{\gamma}_\textrm{A}+\overline{\gamma}_\textrm{A}\right)}{x}\right)\left(\frac{\sigma_\textrm{A}^{2}}{y}+\frac{\sigma_\textrm{B}^{2} \overline{\gamma}_\textrm{B}}{x}\right)}$. In Section IV, we will validate the derived expressions of the ESRs of the four schemes and compare them via numerical experiments.

\subsection{DMT Performance}
The DMT is a fundamental metric to characterize the tradeoff between the reliability and the effectiveness of communication systems \cite{Zheng03TIT,Laneman04TIT,Tse04TIT}. Specifically, the reliability is measured by the diversity gain, which can be defined as the decaying rate of the outage probability with an increase of the SNR in the high SNR region. In this paper, it can be written as
\begin{align}
d=\lim _{P_{\textrm{p}} \rightarrow \infty} \frac{-\log _{2}\left(P_{\textrm{out}}\right)}{\log _{2}\left(P_{\textrm{p}}\right)}.
\label{E20_DiversityGain}
\end{align}
On the other hand, the effectiveness is measured in terms of the multiplexing gain, which is defined as the ratio of the target data rate to the maximum achievable data rate (a.k.a. the increasing rate of the target data rate with an increase of the SNR) in the high SNR region. For user A and user B, it can be represented, respectively, as
\begin{align}
r_{\textrm{A}}=\lim _{P_{\textrm{p}} \rightarrow \infty} \frac{R_{\textrm{A}}}{\log _{2}\left(1+\frac{P_{\textrm{p}} \lambda_{\textrm{A}}}{\sigma_{\textrm{A}}^{2}}\right)},
\label{E21_MultiplexingGainA}
\end{align}
\begin{align}
r_{\textrm{B}}=\lim _{P_{\textrm{p}} \rightarrow \infty} \frac{R_{\textrm{B}}}{\log _{2}\left(1+\frac{P_{\textrm{p}} \lambda_{\textrm{B}}}{\sigma_{\textrm{B}}^{2}}\right)}.
\label{E22_MultiplexingGainB}
\end{align}
To achieve a higher multiplexing gain, the diversity gain would be impaired since the outage probability is generally an increasing function of the target data rate. Therefore, a question is what diversity gain $d(r_{\textrm{A}},r_{\textrm{B}})$ can be achieved for given multiplexing gains $r_{\textrm{A}}$ and $r_{\textrm{B}}$, which is known as the DMT analysis. Based on the derived expressions of the SOP in Appendix B, the DMT performance of the four schemes are investigated and the results are summarized in the following proposition.

\textbf{\emph{Proposition 4}}: The DMT performance of the NC-NOMA scheme and that of the CR-NOMA scheme are $d(r_{\textrm{A}},r_{\textrm{B}})=\min\left\lbrace1-r_{\textrm{A}},1-r_{\textrm{B}},2-2r_{\textrm{A}}-2r_{\textrm{B}}\right\rbrace $ and $d(r_{\textrm{A}},r_{\textrm{B}})=(2-4r_{\textrm{A}}-4r_{\textrm{B}})$, respectively, whereas for both the IR-NOMA scheme and the BC-NOMA scheme, the DMT performance is $d(r_{\textrm{A}},r_{\textrm{B}})=(2-2r_{\textrm{A}}-2r_{\textrm{B}})$.

\emph{Proof}: Please refer to Appendix C.\hfill$\blacksquare$

\emph{Remark 3}: It follows from Proposition 4 that the maximum achievable diversity gain\footnote{The maximum achievable diversity gain (a.k.a. diversity order) denotes the total number of random fading coefficients that a scheme can average over, which is achieved by fixing the target data rates (i.e., zero multiplexing gains).} of the NC-NOMA scheme is merely 1, whereas the counterparts of the three cooperative schemes are 2. This is because for the three cooperative schemes, one more link (i.e., the link between the two users) is constructed to enhance the reception at the user with a worse instantaneous channel condition (i.e., user 2), which helps combat the fading of the direct links.

\emph{Remark 4}: It follows from Proposition 4 that the maximum achievable multiplexing gain\footnote{When the operating multiplexing gain equals the maximum achievable multiplexing gain, there is no protection against the fading of channels (i.e., zero diversity gain), and the outage probability will not decrease with the increase of SNR any more.} of the CR-NOMA scheme is 1/2. Specifically, the sum of $r_{\textrm{A}}$ and $r_{\textrm{B}}$ should be less than 1/2, but each of them can approach to 1/2 with another equal to zero. In comparison, the maximum achievable multiplexing gains of all the other three schemes are 1. This is due to the fact that an extra cooperative transmission phase is needed for user 1 to assist user 2 for the CR-NOMA scheme, which impairs the data rate. Note that the IR-NOMA scheme also involves the cooperative transmission phase but has a maximum achievable multiplexing gain of 1. This is because for the IR-NOMA scheme, the cooperative transmission phase is introduced only when it is necessary to avoid information outage, and in the high SNR region, the cooperative transmission phase is rarely necessary since both users are able to recover their information from the direct transmission in most fading blocks.
\section{Numerical Results and Discussion}
In this section, representative numerical results are provided to validate the theoretical analyses in Section III, and to compare the performance of the four schemes under Rayleigh fading channels. Without loss of generality, we set $\lambda_{\textrm{A}}=1$, $\lambda_{\textrm{B}}=0.5$, $\lambda_{\textrm{g}}=0.5$, $R_{\textrm{A}}=1$bit/s/Hz, $R_{\textrm{B}}=0.5$bit/s/Hz, and $\eta=0.5$ in simulations unless otherwise specified. The ratio of the maximum allowed total power (i.e., $P_\textrm{p}$) to the normalized noise variance (i.e., $\sigma_{\textrm{A}}^2=\sigma_{\textrm{B}}^2=1$) in dB is used to measure the strength of the total transmit power in the rest of the paper.

\begin{figure*}[htp]
	\centering
	\begin{minipage}[t]{0.49\textwidth}
		\centering
		\includegraphics[scale=0.6]{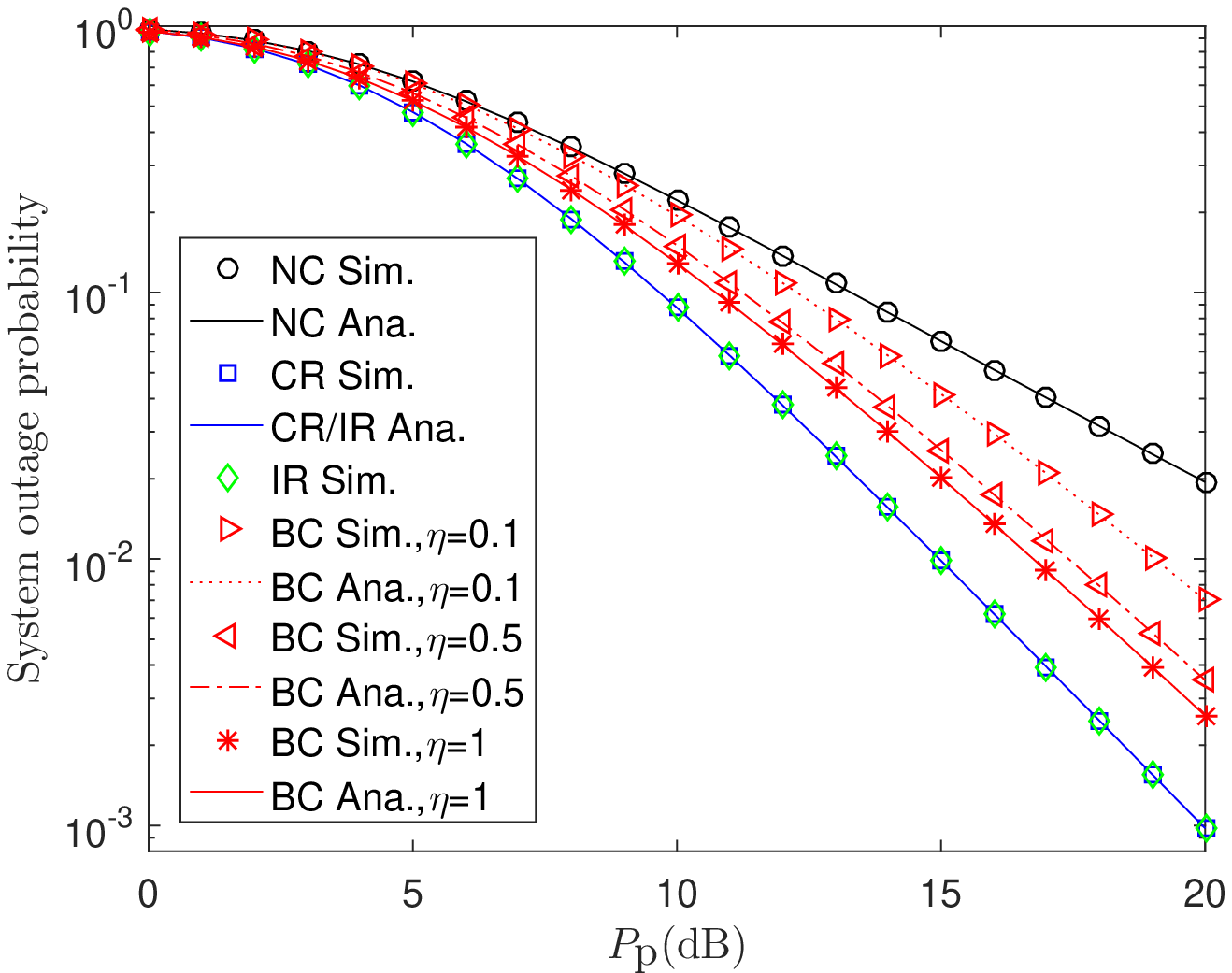}
		\caption{Comparison of the SOP.}
		\label{Fig2SOP}
	\end{minipage}
	\hfil
	\begin{minipage}[t]{0.49\textwidth}
		\includegraphics[scale=0.6]{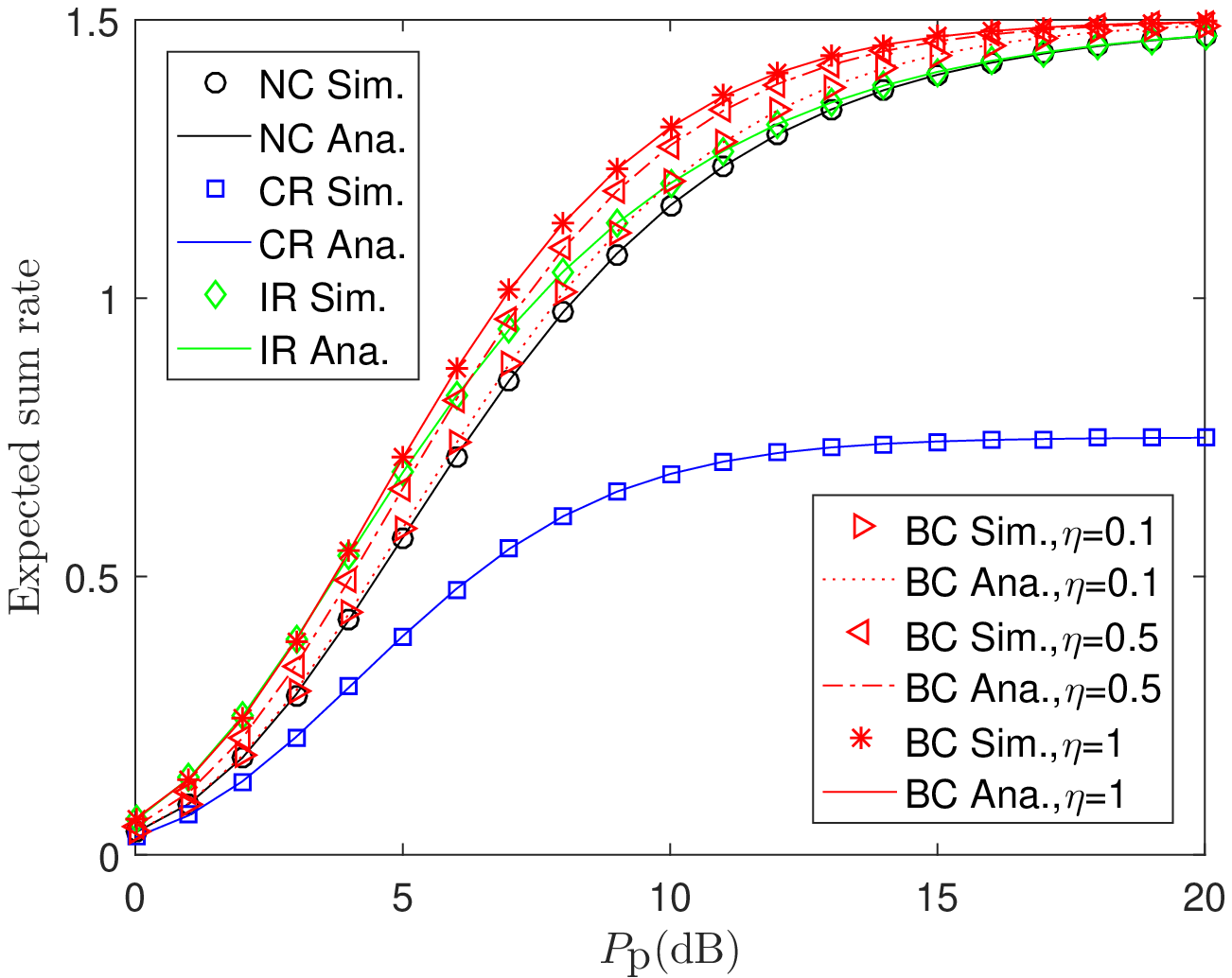}
		\caption{Comparison of the ESR.}
		\label{Fig3ESP}
	\end{minipage}
	\hfil
\end{figure*}
Fig. \ref{Fig2SOP} presents the SOPs of the four schemes, which shows that the simulation results match well with the analytical results. Particularly, it is shown that the SOPs of the three cooperative schemes (i.e., CR, IR, and BC) are lower than that of NC-NOMA over the whole SNR region, which agrees with Corollary 1. Furthermore, it can be observed that CR/IR-NOMA has a lower SOP compared with BC-NOMA. One reason is that user 1 only needs to forward $x_2$ to user 2 for CR/IR-NOMA, whereas for BC-NOMA, both $x_1$ and $x_2$ are backscattered by user 1, which means that the backscattered signals contain both useful signals and interference. Another reason is that the cooperative signals in CR/IR-NOMA are directly transmitted from user 1 to user 2. In comparison, the cooperative signals in BC-NOMA are transmitted from the BS and then backscattered by user 1, and thus suffer more from the path loss. Nonetheless, unlike CR/IR-NOMA, the backscatter cooperation carried out at user 1 does not involve generating carrier signals, which is more energy-efficient and low-cost in practice. 

Fig. \ref{Fig3ESP} demonstrates the ESRs of the four schemes, which shows that the simulation results match well with the analytical results. From the figure, several observations can be drawn: 1) Regardless of the value of the backscatter efficiency, BC-NOMA outperforms NC-NOMA in terms of the ESR over the whole SNR region, which agrees with Corollary 2; 2) With a perfect backscatter efficiency ($\eta=1$), the proposed BC-NOMA outperforms the other three schemes in terms of the ESR, which demonstrates its high effectiveness. However, when the backscatter efficiency is less than one, it is shown that its ESR is lower than that of IR-NOMA in the low SNR region; 3) CR-NOMA has the worst ESR. This is due to the fact that it incorporates the cooperative transmission phase even if it is not necessary, which improves the reliability at the cost of the effectiveness; 4) As the SNR approaches to infinity, the limiting ESR of CR-NOMA is half of the counterparts of the other three schemes. This is intuitive according to the definition of the ESR in \eqref{E18_RES}, since both $P_{\textrm{out}}$ and $P_{\textrm{CT}}$ of the other three schemes approach to zero as the SNR approaches to infinity, whereas the term $P_{\textrm{CT}}$ of CR-NOMA is always one. 5) In the high SNR region, the ESR of IR-NOMA is almost the same as that of NC-NOMA. This is because in the high SNR region, the direct transmission phase is sufficient for both users to recover their information in most fading blocks, in which case IR-NOMA rarely activates the cooperative transmission phase and thus works like NC-NOMA.

\begin{figure*}[htp]
	\centering
	\begin{minipage}[t]{0.45\textwidth}
		\centering
		\includegraphics[scale=0.6]{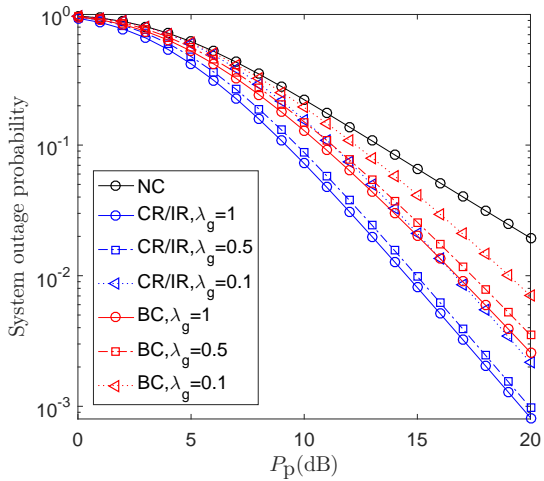}
		\caption{Effects of the average channel condition of the cooperative channel on SOP.}
		\label{Fig4SOP_g}
	\end{minipage}
	\hfil
	\begin{minipage}[t]{0.54\textwidth}
		\includegraphics[scale=0.6]{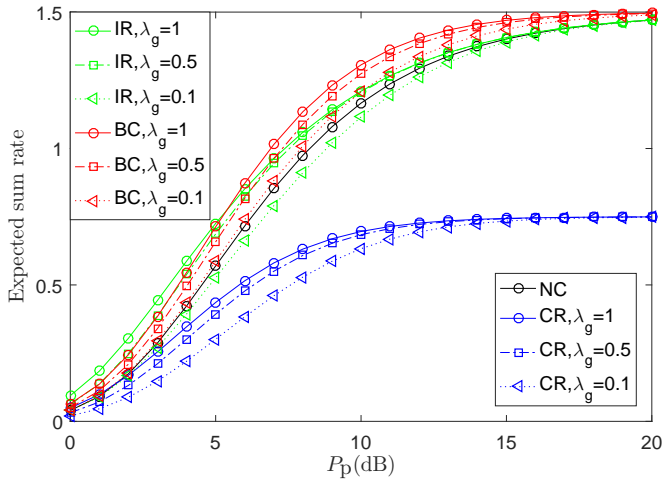}
		\caption{Effects of the average channel condition of the cooperative channel on ESR.}
		\label{Fig5ESR_g}
	\end{minipage}
	\hfil
\end{figure*}
Fig. \ref{Fig4SOP_g} presents the influences of the average channel power gain of the cooperative channel (i.e., $\lambda_{\textrm{g}}$) on the SOP for the four schemes. From the figure, it can be observed that regardless of the (non-zero) value of $\lambda_{\textrm{g}}$, the SOPs of the three cooperative schemes (i.e., CR, IR, and BC) are lower than that of NC-NOMA over the whole SNR region, which again validates Corollary 1. In a similar way, Fig. \ref{Fig5ESR_g} shows the influences of the value of $\lambda_{\textrm{g}}$ on the ESR for the four schemes. An interesting observation is that the ESR of IR-NOMA is lower than that of NC-NOMA when $\lambda_{\textrm{g}}$ is small (e.g., $\lambda_{\textrm{g}}=0.1$). In comparison, regardless of the (non-zero) value of $\lambda_{\textrm{g}}$, the proposed BC-NOMA can always improve the ESR over NC-NOMA. This is because compared with IR-NOMA, which reduces the minimum required total power at the cost of introducing extra time slots, BC-NOMA improves the reliability without impairing the data rate.

\begin{figure}[htp]
	\centering
	\includegraphics[scale=0.6]{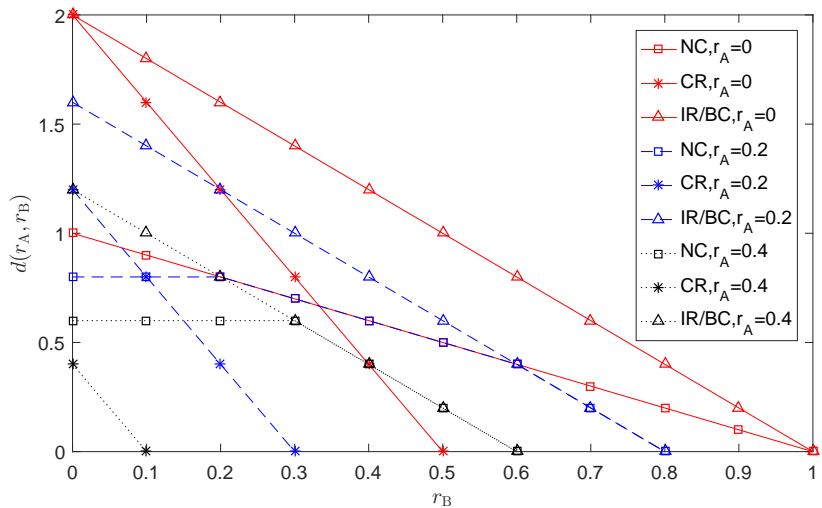}
	\caption{The DMT performance of the four schemes.}
	\label{Fig6DMT}
\end{figure}
Fig. \ref{Fig6DMT} illustrates the DMT performance of the four schemes under different given multiplexing gains at user A\footnote{Note that user A and user B are peers so that the DMT performance of the four schemes under different given multiplexing gains at user B is similar and thus is not illustrated herein.}. From the figure, we can observe that the achievable diversity gain of CR-NOMA is higher than that of NC-NOMA when the given multiplexing gains are small, whereas the achievable diversity gain of CR-NOMA is lower than that of NC-NOMA when the multiplexing gains are large. This is because CR-NOMA can enhance the reliability via user cooperation, but compared with NC-NOMA, CR-NOMA has to sacrifice more reliability to enhance the multiplexing gains as a result of the fact that it introduces a cooperative transmission phase after each direct transmission phase, which impairs the data rate. On the other hand, it is shown that when the given multiplexing gains are high, the achievable diversity gain of IR/BC-NOMA is the same as that of NC-NOMA, whereas IR/BC-NOMA can improve the achievable diversity gain when the multiplexing gains are small. This improvement in transmission reliability benefits from the cooperative behaviors.
\section{Concluding Remarks}
In this paper, a BC-NOMA scheme was proposed, where the user with a better instantaneous channel condition splits and then backscatters part of its received signals to produce constructive multi-path signals to improve the reception at the user with a worse instantaneous channel condition. For comparison, the NC-NOMA scheme, the CR-NOMA scheme, and the IR-NOMA scheme were introduced. The close-form expressions of the minimum required total power to avoid information outage were obtained for all the four schemes, which showed the three cooperative schemes indeed reduce the minimum required total power compared with the NC-NOMA scheme.

Furthermore, we developed the SOPs, ESRs, and the DMT performance of the four schemes under Rayleigh fading channels, which showed that the proposed BC-NOMA scheme strictly outperforms the NC-NOMA scheme in terms of all the three metrics. Finally, representative numerical results were presented to validate the theoretical results, which showed that only the BC-NOMA scheme can enhance the reliability without impairing the effectiveness. This observation demonstrates the benefits of applying the backscatter technique as an alternative to the conventional relaying operation.

\appendices

\section{}
\setcounter{equation}{0}
\renewcommand{\theequation}{\thesection.\arabic{equation}}
\noindent \textbf{A-1: Proof of Proposition 2}

Constraints $\gamma_{12}\ge \overline{\gamma}_{2}$ and $\gamma_{22,\textrm{MRC}}\ge \overline{\gamma}_{2}$ can be combined and rewritten as
\begin{align}
P_{2} \geq P_{1} \overline{\gamma}_{2}+\max \left\{\frac{\sigma_{1}^{2} \overline{\gamma}_{2}}{\left|h_{1}\right|^{2}}, \frac{\sigma_{2}^{2} \overline{\gamma}_{2}}{\left|h_{2}\right|^{2}}-\left(\frac{1}{\left|h_{2}\right|^{2}}+\frac{P_{1}}{\sigma_{2}^{2}}\right) P_{\textrm{h}}|g|^{2}\right\}.
\label{A1_P2min}
\end{align}
Note that the right hand side (RHS) of \eqref{A1_P2min} is a monotonically increasing function of $P_1$. Combining this observation with the fact that $\gamma_{11}\ge \overline{\gamma}_{1}$ can be rewritten as $P_1\ge \frac{\sigma_{1}^{2} \overline{\gamma}_{1}}{\left|h_{1}\right|^{2}}$, we have
\begin{align}
P_{1}^{*}=\frac{\sigma_{1}^{2} \overline{\gamma}_{1}}{\left|h_{1}\right|^{2}}.
\label{A2_P1opt}
\end{align}
Next, by utilizing \eqref{A1_P2min}, P2 can be rewritten as the following problem.
\begin{align}
\textrm{P2a}: \min _{P_{\textrm{h}}} P_{1}^{*}+P_{1}^{*} \overline{\gamma}_{2}+
\underbrace{\max \left\{\frac{\sigma_{1}^{2} \overline{\gamma}_{2}}{\left|h_{1}\right|^{2}}, \frac{\sigma_{2}^{2} \overline{\gamma}_{2}}{\left|h_{2}\right|^{2}}-\left(\frac{1}{\left|h_{2}\right|^{2}}+\frac{P_{1}^{*}}{\sigma_{2}^{2}}\right) P_{\textrm{h}}|g|^{2}\right\}+P_{\textrm{h}}}_{\phi (P_h)}.
\label{A3_P2a}
\end{align}
Furthermore, by making use of \eqref{A2_P1opt}, our goal becomes to choose the optimal $P_\textrm{h}$ to minimize
\begin{align}
\phi\left(P_{\textrm{h}}\right)=\max \left\{\frac{\sigma_{1}^{2} \overline{\gamma}_{2}}{\left|h_{1}\right|^{2}}+P_{\textrm{h}}, \frac{\sigma_{2}^{2} \overline{\gamma}_{2}}{\left|h_{2}\right|^{2}}+\left(1-\left(\frac{1}{\left|h_{2}\right|^{2}}+\frac{\sigma_{1}^{2} \overline{\gamma}_{1}}{\sigma_{2}^{2}\left|h_{1}\right|^{2}}\right)|g|^{2}\right) P_{\textrm{h}}\right\}.
\label{A4_phiPh}
\end{align}
Note that when $|g|^{2} \leq\left(\frac{1}{\left|h_{2}\right|^{2}}+\frac{\sigma_{1}^{2} \overline{\gamma}_{1}}{\sigma_{2}^{2}\left|h_{1}\right|^{2}}\right)^{-1}$, both terms of the RHS of \eqref{A4_phiPh} are monotonically increasing functions of $P_\textrm{h}$. Therefore, in this case, we have $P_\textrm{h}^*=0$ and $\phi\left(P_{\textrm{h}}^{*}\right)=\frac{\sigma_{2}^{2} \overline{\gamma}_{2}}{\left|h_{2}\right|^{2}}$. On the other hand, when $|g|^{2} > \left(\frac{1}{\left|h_{2}\right|^{2}}+\frac{\sigma_{1}^{2} \overline{\gamma}_{1}}{\sigma_{2}^{2}\left|h_{1}\right|^{2}}\right)^{-1}$, the second term of the RHS of \eqref{A4_phiPh} becomes a monotonically decreasing function of $P_\textrm{h}$. In this case, the optimal $P_\textrm{h}$ should make the two terms equal. One can readily show that the corresponding solution is $P_{\textrm{h}}^{*}=\left(\frac{\sigma_{2}^{2} \overline{\gamma}_{2}}{\left|h_{2}\right|^{2}}-\frac{\sigma_{1}^{2} \overline{\gamma}_{2}}{\left|h_{1}\right|^{2}}\right)\left(\frac{\sigma_{1}^{2} \overline{\gamma}_{1}}{\sigma_{2}^{2}\left|h_{1}\right|^{2}}+\frac{1}{\left|h_{2}\right|^{2}}\right)^{-1}|g|^{-2}$. Finally, the proof can be completed by inserting $P_{\textrm{h}}^{*}$ into \eqref{A3_P2a}.

\noindent \textbf{A-2: Proof of Proposition 3}

Constraints $\gamma_{12,\textrm{BT}}\ge \overline{\gamma}_{2}$ and $\gamma_{22,\textrm{BT}}\ge \overline{\gamma}_{2}$ can be combined and rewritten as
\begin{align}
P_{2} \geq P_{1} \overline{\gamma}_{2}+\max \left\{\frac{\sigma_{1}^{2} \overline{\gamma}_{2}}{\left|h_{1}\right|^{2}\left(1-\beta_{1}\right)}, \frac{\sigma_{2}^{2} \overline{\gamma}_{2}}{\left|h_{2}\right|^{2}+\beta_{1}\left|h_{1}\right|^{2}|g|^{2}}\right\}.
\label{A5_P2min2}
\end{align}
Note that the RHS of \eqref{A5_P2min2} is a monotonically increasing function of $P_1$. Combining this observation with the fact that $\gamma_{11,\textrm{BT}}\ge \overline{\gamma}_{1}$ can be rewritten as $P_{1} \geq \frac{\sigma_{1}^{2} \overline{\gamma}_{1}}{\left|h_{1}\right|^{2}\left(1-\beta_{1}\right)}$, for any given $\beta_{1}$, we have
\begin{align}
P_{1}^{*}=\frac{\sigma_{1}^{2} \overline{\gamma}_{1}}{\left|h_{1}\right|^{2}\left(1-\beta_{1}\right)}.
\label{A6_P1opt2}
\end{align}
Next, by utilizing \eqref{A5_P2min2} and \eqref{A6_P1opt2}, P3 can be rewritten as the following problem.
\begin{align}
\textrm{P3a}: \min _{\beta_{1}}
\underbrace{\frac{\sigma_{1}^{2}\left(\overline{\gamma}_{1}+\overline{\gamma}_{1} \overline{\gamma}_{2}\right)}{\left|h_{1}\right|^{2}\left(1-\beta_{1}\right)}+\max \left\{\frac{\sigma_{1}^{2} \overline{\gamma}_{2}}{\left|h_{1}\right|^{2}\left(1-\beta_{1}\right)}, \frac{\sigma_{2}^{2} \overline{\gamma}_{2}}{\left|h_{2}\right|^{2}+\beta_{1}\left|h_{1}\right|^{2}|g|^{2}}\right\}}_{\psi(\beta_{1})}.
\label{A7_P3a}
\end{align}

Hereafter, our goal becomes to choose the optimal $\beta_{1}$ to minimize $\psi(\beta_{1})$. Before that, we need to compare the two terms in $\max \left\{\frac{\sigma_{1}^{2} \overline{\gamma}_{2}}{\left|h_{1}\right|^{2}\left(1-\beta_{1}\right)}, \frac{\sigma_{2}^{2} \overline{\gamma}_{2}}{\left|h_{2}\right|^{2}+\beta_{1}\left|h_{1}\right|^{2}|g|^{2}}\right\}$. When $\beta_{1}> \frac{\sigma_{2}^{2}\left|h_{1}\right|^{2}-\sigma_{1}^{2}\left|h_{2}\right|^{2}}{\sigma_{2}^{2}\left|h_{1}\right|^{2}+\sigma_{1}^{2}\left|h_{1}\right|^{2}|g|^{2}} \triangleq \widehat{\beta_{1}}$, the first term is larger and we have $\psi(\beta_{1})=\frac{\sigma_{1}^{2}(\overline{\gamma}_{1}+\overline{\gamma}_{1}\overline{\gamma}_{2}+\overline{\gamma}_{2})}{\left|h_{1}\right|^{2}\left(1-\beta_{1}\right)}$, which is a monotonically increasing function of $\beta_{1}$.

On the other hand, when $\beta_{1}\le\widehat{\beta_{1}}$, we have $\psi(\beta_{1})=\frac{\sigma_{1}^{2}(\overline{\gamma}_{1}+\overline{\gamma}_{1}\overline{\gamma}_{2})}{\left|h_{1}\right|^{2}\left(1-\beta_{1}\right)}+\frac{\sigma_{2}^{2} \overline{\gamma}_{2}}{\left|h_{2}\right|^{2}+\beta_{1}\left|h_{1}\right|^{2}|g|^{2}}$, whose derived function is $\psi^{\prime}(\beta_{1})=\frac{\sigma_{1}^{2}(\overline{\gamma}_{1}+\overline{\gamma}_{1}\overline{\gamma}_{2})}{\left|h_{1}\right|^{2}\left(1-\beta_{1}\right)^{2}}-\frac{\sigma_{2}^{2} \overline{\gamma}_{2}\left|h_{1}\right|^{2}|g|^{2}}{\left(\left|h_{2}\right|^{2}+\beta_{1}\left|h_{1}\right|^{2}|g|^{2}\right)^{2}}$. Furthermore, one can show that $\psi^{\prime}(\beta_{1})$ equals zero when $\beta_{1}=\left(\sqrt{\frac{\sigma_{2}^{2} \overline{\gamma}_{2}|g|^{2}}{\sigma_{1}^{2}(\overline{\gamma}_{1}+\overline{\gamma}_{1}\overline{\gamma}_{2})}}-\frac{\left|h_{2}\right|^{2}}{\left|h_{1}\right|^{2}}\right)\left(\sqrt{\frac{\sigma_{2}^{2} \overline{\gamma}_{2}|g|^{2}}{\sigma_{1}^{2}(\overline{\gamma}_{1}+\overline{\gamma}_{1}\overline{\gamma}_{2})}}+|g|^{2}\right)^{-1}\triangleq\overline{\beta_{1}}$. Note that if $0<\overline{\beta_{1}}<\widehat{\beta_{1}}$, as the increase of $\beta_{1}$, $\psi(\beta_{1})$ first decreases within $[0,\overline{\beta_{1}})$ and then increases within $(\overline{\beta_{1}},\widehat{\beta_{1}}]$. If $\overline{\beta_{1}}\le0$, $\psi(\beta_{1})$ is a monotonically increasing function of $\beta_{1}$ within $[0,\widehat{\beta_{1}}]$, whereas if $\overline{\beta_{1}}\ge\widehat{\beta_{1}}$, $\psi(\beta_{1})$ is a monotonically decreasing function of $\beta_{1}$ within $[0,\widehat{\beta_{1}}]$.

By combining the two cases discussed above, we have
\begin{align}
\beta_{1}^{*}=
\begin{cases}
0,&\overline{\beta_{1}}\le 0,\\
\overline{\beta_{1}},&0<\overline{\beta_{1}}<\widehat{\beta_{1}},\\
\widehat{\beta_{1}},&\overline{\beta_{1}} \geq \widehat{\beta_{1}}.
\end{cases}
\label{A8_betaopt}
\end{align}
By inserting \eqref{A8_betaopt} into \eqref{A7_P3a}, we complete the proof.

\section{Derivation of the SOP}
\setcounter{equation}{0}
Note that the first term of the RHS of \eqref{E17_Pout} can be written as
\begin{align}
P_{\textrm{out}}^{\prime}=\operatorname{Pr}\left(P_{\textrm{min}}>P_{\textrm{p}}, \frac{\left|h_{1}\right|^{2}}{\sigma_{1}^{2}} \geq \frac{\left|h_{2}\right|^{2}}{\sigma_{2}^{2}}\right),
\label{B1_Pout'}
\end{align}
where $P_1\triangleq P_{\textrm{A}}$, $P_2\triangleq P_{\textrm{B}}$, $\sigma_{1}^2\triangleq \sigma_{\textrm{A}}^2$, $\sigma_{2}^2\triangleq \sigma_{\textrm{B}}^2$, $n_1\triangleq n_{\textrm{A}}$, $n_2\triangleq n_{\textrm{B}}$, $R_1\triangleq R_{\textrm{A}}$, $R_2\triangleq R_{\textrm{B}}$, $h_1\triangleq h_{\textrm{A}}$, and $h_2\triangleq h_{\textrm{B}}$. Similarly, the second term of \eqref{E17_Pout} can also be written as \eqref{B1_Pout'}, where $P_1\triangleq P_{\textrm{B}}$, $P_2\triangleq P_{\textrm{A}}$, $\sigma_{1}^2\triangleq \sigma_{\textrm{B}}^2$, $\sigma_{2}^2\triangleq \sigma_{\textrm{A}}^2$, $n_1\triangleq n_{\textrm{B}}$, $n_2\triangleq n_{\textrm{A}}$, $R_1\triangleq R_{\textrm{B}}$, $R_2\triangleq R_{\textrm{A}}$, $h_1\triangleq h_{\textrm{B}}$, and $h_2\triangleq h_{\textrm{A}}$. This means that hereafter we only need to develop the analytical expression of \eqref{B1_Pout'}.

\noindent \textbf{B-1: The SOP of the NC-NOMA Scheme}

In the rest of the paper, we define $\lambda_{x}$, $\lambda_{y}$, and $\lambda_{z}$ as the means of $x\triangleq \left|h_{1}\right|^{2}$, $y\triangleq \left|h_{2}\right|^{2}$, and $z\triangleq \left|g\right|^{2}$, respectively. According to the expression of $P_{\textrm{min,NC}}$ in Proposition 1, for the NC-NOMA scheme, \eqref{B1_Pout'} can be rewritten as
\begin{align}
P_{\textrm{out,NC}}^{\prime}=\operatorname{Pr}\left(\frac{\sigma_{1}^{2} (\overline{\gamma}_{1}+\overline{\gamma}_{1} \overline{\gamma}_{2})}{x}+\frac{\sigma_{2}^{2} \overline{\gamma}_{2}}{y}>P_{\textrm{p}},\frac{x}{\sigma_{1}^{2}} \geq \frac{y}{\sigma_{2}^{2}}\right).
\label{B2_Pout'NC}
\end{align}
Next, we divide \eqref{B2_Pout'NC} into $P_{\textrm{out,NC}}^{\prime}=P_{\textrm{out,NC1}}^{\prime}+P_{\textrm{out,NC2}}^{\prime}$, where $P_{\textrm{out,NC1}}^{\prime}\triangleq \operatorname{Pr}\left(y \leq \frac{\sigma_{2}^{2} \overline{\gamma}_{2}}{P_{\textrm{p}}}, x \geq \frac{\sigma_{1}^{2}}{\sigma_{2}^{2}} y\right)$ and $P_{\textrm{out,NC2}}^{\prime}\triangleq \operatorname{Pr}\left(\frac{\sigma_{1}^{2}}{\sigma_{2}^{2}} y \leq x<\frac{\sigma_{1}^{2} (\overline{\gamma}_{1}+\overline{\gamma}_{1} \overline{\gamma}_{2})}{P_{\textrm{p}}-\frac{\sigma_{2}^{2} \overline{\gamma}_{2}}{y}}, \frac{\sigma_{2}^{2} \overline{\gamma}_{2}}{P_{\textrm{p}}}<y<\frac{\sigma_{2}^{2}(\overline{\gamma}_{1}+\overline{\gamma}_{1} \overline{\gamma}_{2}+\overline{\gamma}_{2})}{P_{\textrm{p}}}\right)$. Furthermore, we have
\begin{align}
P_{\textrm{out,NC1}}^{\prime}=\int_{0}^{\frac{\sigma_{2}^{2} \overline{\gamma}_{2}}{P_{\textrm{p}}}} \int_{\frac{\sigma_{1}^{2}}{\sigma_{2}^{2}} y}^{\infty} \frac{e^{-\frac{x}{\lambda_{x}}}}{\lambda_{x}} dx \frac{e^{-\frac{y}{\lambda_{y}}}}{\lambda_{y}} dy=\frac{1-e^{-\left(\frac{\sigma_{1}^{2}}{\lambda_{x} \sigma_{2}^{2}}+\frac{1}{\lambda_{y}}\right) \frac{\sigma_{2}^{2} \overline{\gamma}_{2}}{P_{\textrm{p}}}}}{\lambda_{y}\left(\frac{\sigma_{1}^{2}}{\lambda_{x} \sigma_{2}^{2}}+\frac{1}{\lambda_{y}}\right)}.
\label{B3_Pout'NC1}
\end{align}
In the same way, one can show that
\begin{align}
P_{\textrm{out,NC2}}^{\prime}=\int_{\frac{\sigma_{2}^{2} \overline{\gamma}_{2}}{P_{\textrm{p}}}}^{\frac{\sigma_{2}^{2}(\overline{\gamma}_{1}+\overline{\gamma}_{1} \overline{\gamma}_{2}+\overline{\gamma}_{2})}{P_{\textrm{p}}}}
\left( e^{-\frac{\sigma_{1}^{2}}{\lambda_{x} \sigma_{2}^{2}} y}-
e^{-\frac{\sigma_{1}^{2} (\overline{\gamma}_{1}+\overline{\gamma}_{1} \overline{\gamma}_{2})}{\lambda_{x}\left(P_{\textrm{p}}-\frac{\sigma_{2}^{2} \overline{\gamma}_{2}}{y}\right)}}\right) \frac{e^{-\frac{y}{\lambda_{y}}}}{\lambda_{y}} dy.
\label{B4_Pout'NC2}
\end{align}
Unfortunately, a more concise form of $P_{\textrm{out,NC2}}^{\prime}$ cannot be achieved due to the complicated integral. Combining the foregoing results, we complete the derivation of the SOP for NC-NOMA.

\noindent \textbf{B-2: The SOP of the CR/IR-NOMA Scheme}

According to Proposition 2, for the CR/IR-NOMA scheme, \eqref{B1_Pout'} can be rewritten as
\begin{align}
P_{\textrm{out,CR/IR}}^{\prime}=&\operatorname{Pr}\left(\frac{x}{\sigma_{1}^{2}} \geq \frac{y}{\sigma_{2}^{2}}, z \leq\left(\frac{1}{y}+\frac{\sigma_{1}^{2} \overline{\gamma}_{1}}{\sigma_{2}^{2} x}\right)^{-1}, \frac{\sigma_{1}^{2}\left(\overline{\gamma}_{1}+\overline{\gamma}_{1} \overline{\gamma}_{2}\right)}{x}+\frac{\sigma_{2}^{2} \overline{\gamma}_{2}}{y}>P_{\textrm{p}}\right)\nonumber\\
&+\operatorname{Pr}\left(\frac{x}{\sigma_{1}^{2}} \geq \frac{y}{\sigma_{2}^{2}}, z>\left(\frac{1}{y}+\frac{\sigma_{1}^{2} \overline{\gamma}_{1}}{\sigma_{2}^{2} x}\right)^{-1}, \frac{\sigma_{1}^{2}\left(\overline{\gamma}_{1}+\overline{\gamma}_{1} \overline{\gamma}_{2}+\overline{\gamma}_{2}\right)}{x}+\frac{\overline{\gamma}_{2}\left(\frac{\sigma_{2}^{2}}{y}-\frac{\sigma_{1}^{2}}{x}\right)}{z\left(\frac{\sigma_{1}^{2} \overline{\gamma}_{1}}{\sigma_{2}^{2} x}+\frac{1}{y}\right)}>P_{\textrm{p}}\right).
\label{B5_Pout'TRIR}
\end{align}
The first term of the RHS of \eqref{B5_Pout'TRIR} can be divided into two parts according to the relative size of $\frac{\sigma_{2}^{2} \overline{\gamma}_{2}}{y}$ and $P_{\textrm{p}}$. They are $Q_{1,1}\triangleq \operatorname{Pr}\left(\frac{x}{\sigma_{1}^{2}} \geq \frac{y}{\sigma_{2}^{2}}, z \leq\left(\frac{1}{y}+\frac{\sigma_{1}^{2} \overline{\gamma}_{1}}{\sigma_{2}^{2} x}\right)^{-1},\frac{\sigma_{2}^{2} \overline{\gamma}_{2}}{y}\ge P_{\textrm{p}}\right)$ and $Q_{1,2}\triangleq \operatorname{Pr}\left(\frac{x}{\sigma_{1}^{2}} \geq \frac{y}{\sigma_{2}^{2}}, z \leq\left(\frac{1}{y}+\frac{\sigma_{1}^{2} \overline{\gamma}_{1}}{\sigma_{2}^{2} x}\right)^{-1},\frac{\sigma_{2}^{2} \overline{\gamma}_{2}}{y}<P_{\textrm{p}},
\frac{\sigma_{1}^{2}\left(\overline{\gamma}_{1}+\overline{\gamma}_{1} \overline{\gamma}_{2}\right)}{x}+\frac{\sigma_{2}^{2} \overline{\gamma}_{2}}{y}>P_{\textrm{p}}\right)$. Similarly, the second term of the RHS of \eqref{B5_Pout'TRIR} can be divided into two parts according to the relative size of $\frac{\sigma_{1}^{2}\left(\overline{\gamma}_{1}+\overline{\gamma}_{1} \overline{\gamma}_{2}+\overline{\gamma}_{2}\right)}{x}$ and $P_{\textrm{p}}$. They are $Q_{2,1}\triangleq \operatorname{Pr}\left(\frac{x}{\sigma_{1}^{2}} \geq \frac{y}{\sigma_{2}^{2}}, z>\left(\frac{1}{y}+\frac{\sigma_{1}^{2} \overline{\gamma}_{1}}{\sigma_{2}^{2} x}\right)^{-1}, \frac{\sigma_{1}^{2}\left(\overline{\gamma}_{1}+\overline{\gamma}_{1} \overline{\gamma}_{2}+\overline{\gamma}_{2}\right)}{x}\ge P_{\textrm{p}}\right)$ and $Q_{2,2}\triangleq \operatorname{Pr}\left(\frac{x}{\sigma_{1}^{2}} \geq \frac{y}{\sigma_{2}^{2}}, z>\left(\frac{1}{y}+\frac{\sigma_{1}^{2} \overline{\gamma}_{1}}{\sigma_{2}^{2} x}\right)^{-1}, \frac{\sigma_{1}^{2}\left(\overline{\gamma}_{1}+\overline{\gamma}_{1} \overline{\gamma}_{2}+\overline{\gamma}_{2}\right)}{x}< P_{\textrm{p}},\frac{\sigma_{1}^{2}\left(\overline{\gamma}_{1}+\overline{\gamma}_{1} \overline{\gamma}_{2}+\overline{\gamma}_{2}\right)}{x}+\frac{\overline{\gamma}_{2}\left(\frac{\sigma_{2}^{2}}{y}-\frac{\sigma_{1}^{2}}{x}\right)}{z\left(\frac{\sigma_{1}^{2} \overline{\gamma}_{1}}{\sigma_{2}^{2} x}+\frac{1}{y}\right)} > P_{\textrm{p}}\right)$. In other words, we have $P_{\textrm{out,CR/IR}}^{\prime}=Q_{1,1}+Q_{1,2}+Q_{2,1}+Q_{2,2}$. In what follows, we determine these four terms one by one. The first term can be given by
\begin{align}
Q_{1,1}=
\int_{0}^{\frac{\sigma_{2}^{2} \overline{\gamma}_{2}}{P_{\textrm{p}}}}
\int_{\frac{\sigma_{1}^{2}}{\sigma_{2}^{2}} y}^{\infty}
\int_{0}^{\left(\frac{1}{y}+\frac{\sigma_{1}^{2} \overline{\gamma}_{1}}{\sigma_{2}^{2} x}\right)^{-1}}
\frac{e^{-\frac{z}{\lambda_{z}}}}{\lambda_{z}} dz
\frac{e^{-\frac{x}{\lambda_{x}}}}{\lambda_{x}} dx
\frac{e^{-\frac{y}{\lambda_{y}}}}{\lambda_{y}} dy.
\label{B6_Q11_1}
\end{align}
After performing some algebraic arrangements and the change of variables, one can show that
\begin{align}
Q_{1,1}=
\frac{\sigma_{1}^{2}}{\lambda_{x} \lambda_{y} \sigma_{2}^{2}} \int_{1}^{\infty}&\left(1-\left(\varphi_{1}(x) \frac{\sigma_{2}^{2} \overline{\gamma}_{2}}{P_{\textrm{p}}}+1\right) e^{-\varphi_{1}(x) \frac{\sigma_{2}^{2} \overline{\gamma}_{2}}{P_{\textrm{p}}}}\right)\left(\varphi_{1}(x)\right)^{-2}
\nonumber\\&-
\left(1-\left(\varphi_{2}(x) \frac{\sigma_{2}^{2} \overline{\gamma}_{2}}{P_{\textrm{p}}}+1\right) e^{-\varphi_{2}(x) \frac{\sigma_{2}^{2} \overline{\gamma}_{2}}{P_{\textrm{p}}}}\right)\left(\varphi_{2}(x)\right)^{-2} dx,
\label{B7_Q11_2}
\end{align}
where we define $\varphi_{1}(x)\triangleq \left(\frac{1}{\lambda_{y}}+\frac{\sigma_{1}^{2} x}{\sigma_{2}^{2} \lambda_{x}}\right)$ and $\varphi_{2}(x)\triangleq \left(\frac{x}{\lambda_{z}(\overline{\gamma}_{1}+x)}+\frac{1}{\lambda_{y}}+\frac{\sigma_{1}^{2} x}{\sigma_{2}^{2} \lambda_{x}}\right)$. Next, after analyzing the integral regions of the remaining three terms, one can show that
\begin{align}
Q_{1,2}&=
\int_{\frac{\sigma_{2}^{2} \overline{\gamma}_{2}}{P_{\textrm{p}}}}^{\frac{\sigma_{2}^{2}(\overline{\gamma}_{1}+\overline{\gamma}_{1}\overline{\gamma}_{2}+\overline{\gamma}_{2})}{P_{\textrm{p}}}}
\int_{\frac{\sigma_{1}^{2}}{\sigma_{2}^{2}} y}^{\frac{\sigma_{1}^{2} (\overline{\gamma}_{1}+\overline{\gamma}_{1} \overline{\gamma}_{2})}{P_{\textrm{p}}-\frac{\sigma_{2}^{2} \overline{\gamma}_{2}}{y}}}
\int_{0}^{\left(\frac{1}{y}+\frac{\sigma_{1}^{2} \overline{\gamma}_{1}}{\sigma_{2}^{2} x}\right)^{-1}}
\frac{e^{-\frac{z}{\lambda_{z}}}}{\lambda_{z}} dz
\frac{e^{-\frac{x}{\lambda_{x}}}}{\lambda_{x}} dx
\frac{e^{-\frac{y}{\lambda_{y}}}}{\lambda_{y}} dy
\nonumber\\&=
\int_{\frac{\sigma_{2}^{2} \overline{\gamma}_{2}}{P_{\textrm{p}}}}^{\frac{\sigma_{2}^{2}(\overline{\gamma}_{1}+\overline{\gamma}_{1}\overline{\gamma}_{2}+\overline{\gamma}_{2})}{P_{\textrm{p}}}}
\int_{\frac{\sigma_{1}^{2}}{\sigma_{2}^{2}} y}^{\frac{\sigma_{1}^{2} (\overline{\gamma}_{1}+\overline{\gamma}_{1} \overline{\gamma}_{2})}{P_{\textrm{p}}-\frac{\sigma_{2}^{2} \overline{\gamma}_{2}}{y}}}
\left(1-e^{-\frac{1}{\lambda_{z}}\left(\frac{1}{y}+\frac{\sigma_{1}^{2} \overline{\gamma}_{1}}{\sigma_{2}^{2} x}\right)^{-1}}\right)
\frac{e^{-\frac{x}{\lambda_{x}}}}{\lambda_{x}} dx
\frac{e^{-\frac{y}{\lambda_{y}}}}{\lambda_{y}} dy,
\label{B8_Q12}
\end{align}
\begin{align}
Q_{2,1}&=
\int_{0}^{\frac{\sigma_{2}^{2}(\overline{\gamma}_{1}+\overline{\gamma}_{1}\overline{\gamma}_{2}+\overline{\gamma}_{2})}{P_{\textrm{p}}}}
\int_{\frac{\sigma_{1}^{2}}{\sigma_{2}^{2}} y}^{\frac{\sigma_{1}^{2}(\overline{\gamma}_{1}+\overline{\gamma}_{1}\overline{\gamma}_{2}+\overline{\gamma}_{2})}{P_{\textrm{p}}}}
\int_{\left(\frac{1}{y}+\frac{\sigma_{1}^{2} \overline{\gamma}_{1}}{\sigma_{2}^{2} x}\right)^{-1}}^{\infty}
\frac{e^{-\frac{z}{\lambda_{z}}}}{\lambda_{z}} dz
\frac{e^{-\frac{x}{\lambda_{x}}}}{\lambda_{x}} dx
\frac{e^{-\frac{y}{\lambda_{y}}}}{\lambda_{y}} dy
\nonumber\\&=
\int_{0}^{\frac{\sigma_{2}^{2}(\overline{\gamma}_{1}+\overline{\gamma}_{1}\overline{\gamma}_{2}+\overline{\gamma}_{2})}{P_{\textrm{p}}}}
\int_{\frac{\sigma_{1}^{2}}{\sigma_{2}^{2}} y}^{\frac{\sigma_{1}^{2}(\overline{\gamma}_{1}+\overline{\gamma}_{1}\overline{\gamma}_{2}+\overline{\gamma}_{2})}{P_{\textrm{p}}}}
e^{-\frac{1}{\lambda_{z}}\left(\frac{1}{y}+\frac{\sigma_{1}^{2} \overline{\gamma}_{1}}{\sigma_{2}^{2} x}\right)^{-1}}
\frac{e^{-\frac{x}{\lambda_{x}}}}{\lambda_{x}} dx
\frac{e^{-\frac{y}{\lambda_{y}}}}{\lambda_{y}} dy,
\label{B9_Q21}
\end{align}
\begin{align}
&Q_{2,2}=
\int_{\frac{\sigma_{1}^{2}(\overline{\gamma}_{1}+\overline{\gamma}_{1}\overline{\gamma}_{2}+\overline{\gamma}_{2})}{P_{\textrm{p}}}}^{\infty}
\int_{0}^{\frac{\sigma_{2}^{2} \overline{\gamma}_{2}}{P_{\textrm{p}}-\frac{\sigma_{1}^{2}(\overline{\gamma}_{1}+\overline{\gamma}_{1}\overline{\gamma}_{2})}{x}}}
\int_{\left(\frac{1}{y}+\frac{\sigma_{1}^{2} \overline{\gamma}_{1}}{\sigma_{2}^{2} x}\right)^{-1}}^{\varphi_{3}(x, y)}
\frac{e^{-\frac{z}{\lambda_{z}}}}{\lambda_{z}} dz
\frac{e^{-\frac{y}{\lambda_{y}}}}{\lambda_{y}} dy
\frac{e^{-\frac{x}{\lambda_{x}}}}{\lambda_{x}} dx
\nonumber\\&=
\int_{\frac{\sigma_{1}^{2}(\overline{\gamma}_{1}+\overline{\gamma}_{1}\overline{\gamma}_{2}+\overline{\gamma}_{2})}{P_{\textrm{p}}}}^{\infty}
\int_{0}^{\frac{\sigma_{2}^{2} \overline{\gamma}_{2}}{P_{\textrm{p}}-\frac{\sigma_{1}^{2}(\overline{\gamma}_{1}+\overline{\gamma}_{1}\overline{\gamma}_{2})}{x}}}
\left( e^{-\frac{1}{\lambda_{z}}\left(\frac{1}{y}+\frac{\sigma_{1}^{2} \overline{\gamma}_{1}}{\sigma_{2}^{2} x}\right)^{-1}}
-e^{-\frac{1}{\lambda_{z}} \varphi_{3}(x, y)}\right) 
\frac{e^{-\frac{y}{\lambda_{y}}}}{\lambda_{y}} dy
\frac{e^{-\frac{x}{\lambda_{x}}}}{\lambda_{x}} dx,
\label{B10_Q22}
\end{align}
where we define $\varphi_{3}(x, y) \triangleq \frac{\overline{\gamma}_{2} \sigma_{2}^{2}\left(\frac{\sigma_{2}^{2}}{y}-\frac{\sigma_{1}^{2}}{x}\right)}{\left(P_{\textrm{p}}-\frac{\sigma_{1}^{2}(\overline{\gamma}_{1}+\overline{\gamma}_{1}\overline{\gamma}_{2}+\overline{\gamma}_{2})}{x}\right)\left(\frac{\sigma_{2}^{2}}{y}+\frac{\sigma_{1}^{2} \overline{\gamma}_{1}}{x}\right)}$. Combining the foregoing results, we complete the derivation of the SOP for the CR/IR-NOMA scheme. More concise expressions cannot be achieved due to the complicated integrals. Nevertheless, the numerical tools at present are efficient enough to determine their values. In addition, the derived expressions above are useful for determining the DMT performance of the CR/IR-NOMA scheme in Appendix C.

\noindent \textbf{B-3: The SOP of the BC-NOMA Scheme}

According to Proposition 3, for the BC-NOMA scheme, \eqref{B1_Pout'} can be rewritten as
\begin{align}
&P_{\textrm{out,BC}}^{\prime}=\operatorname{Pr}\left(z \leq \varphi_{4}(x, y), \frac{x}{\sigma_{1}^{2}} \geq \frac{y}{\sigma_{2}^{2}}, \frac{\sigma_{1}^{2}(\overline{\gamma}_{1}+\overline{\gamma}_{1} \overline{\gamma}_{2})}{x}+\frac{\sigma_{2}^{2} \overline{\gamma}_{2}}{y}>P_{\textrm{p}}\right) 
\nonumber\\&
+\operatorname{Pr}\left(\varphi_{4}(x, y)<z<\varphi_{5}, \frac{x}{\sigma_{1}^{2}} \geq \frac{y}{\sigma_{2}^{2}},\frac{\sigma_{1}^{2}\left(\overline{\gamma}_{1}+\overline{\gamma}_{1} \overline{\gamma}_{2}\right)\left( 2 \sqrt{\frac{\sigma_{2}^{2} \overline{\gamma}_{2} z}{\sigma_{1}^{2}\left( \overline{\gamma}_{1}+\overline{\gamma}_{1} \overline{\gamma}_{2}\right) }}+z\right) +\sigma_{2}^{2} \overline{\gamma}_{2}}{x z+y}>P_{\textrm{p}}\right)
\nonumber\\&
+\operatorname{Pr}\left(z \geq \varphi_{5}, \frac{x}{\sigma_{1}^{2}} \geq \frac{y}{\sigma_{2}^{2}}, \frac{(\overline{\gamma}_{1}+\overline{\gamma}_{1} \overline{\gamma}_{2}+\overline{\gamma}_{2})\left(\sigma_{2}^{2}+\sigma_{1}^{2} z\right)}{x z+y}>P_{\textrm{p}}\right)
\triangleq I_{1}+I_{2}+I_{3},
\label{B11_Pout'BC}
\end{align}
where $\varphi_{4}(x, y) \triangleq\left(\frac{y}{x}\right)^{2} \frac{\sigma_{1}^{2}(\overline{\gamma}_{1}+\overline{\gamma}_{1}\overline{\gamma}_{2})}{\sigma_{2}^{2} \overline{\gamma}_{2}}$ and $\varphi_{5} \triangleq \frac{\sigma_{2}^{2}(\overline{\gamma}_{1}+\overline{\gamma}_{1}\overline{\gamma}_{2})}{\sigma_{1}^{2} \overline{\gamma}_{2}}$. Hereafter, we determine the three terms above one by one. First, after analyzing the integral region, we can rewrite the first term as
\begin{align}
I_{1}=&
\int_{0}^{\frac{\sigma_{1}^{2}(\overline{\gamma}_{1}+\overline{\gamma}_{1}\overline{\gamma}_{2}+\overline{\gamma}_{2})}{P_{\textrm{p}}}}
\int_{0}^{\frac{x \sigma_{2}^{2}}{\sigma_{1}^{2}}}
\int_{0}^{\left(\frac{y}{x}\right)^{2} \frac{\sigma_{1}^{2}(\overline{\gamma}_{1}+\overline{\gamma}_{1}\overline{\gamma}_{2})}{\sigma_{2}^{2} \overline{\gamma}_{2}}}
\frac{e^{-\frac{z}{\lambda_{z}}}}{\lambda_{z}} d z \frac{e^{-\frac{y}{\lambda_{y}}}}{\lambda_{y}} d y \frac{e^{-\frac{x}{\lambda_{x}}}}{\lambda_{x}} d x
\nonumber\\&
+\int_{\frac{\sigma_{1}^{2}(\overline{\gamma}_{1}+\overline{\gamma}_{1}\overline{\gamma}_{2}+\overline{\gamma}_{2})}{P_{\textrm{p}}}}^{\infty}
\int_{0}^{\frac{\sigma_{2}^{2} \overline{\gamma}_{2}}{P_{\textrm{p}}-\frac{\sigma_{1}^{2}(\overline{\gamma}_{1}+\overline{\gamma}_{1}\overline{\gamma}_{2})}{x}}}
\int_{0}^{\left(\frac{y}{x}\right)^{2} \frac{\sigma_{1}^{2}(\overline{\gamma}_{1}+\overline{\gamma}_{1}\overline{\gamma}_{2})}{\sigma_{2}^{2} \overline{\gamma}_{2}}}
\frac{e^{-\frac{z}{\lambda_{z}}}}{\lambda_{z}} d z \frac{e^{-\frac{y}{\lambda_{y}}}}{\lambda_{y}} d y \frac{e^{-\frac{x}{\lambda_{x}}}}{\lambda_{x}} d x
\triangleq I_{1,1}+I_{1,2}.
\label{B12_I1}
\end{align}
Next, by making use of \cite[Eq. (3.322.1)]{TableOfIntegrals} and \cite[Eq. (3.322.2)]{TableOfIntegrals}, one can show that
\begin{align}
I_{1,1}=
\int_{0}^{\frac{\sigma_{1}^{2}(\overline{\gamma}_{1}+\overline{\gamma}_{1}\overline{\gamma}_{2}+\overline{\gamma}_{2})}{P_{\textrm{p}}}}&
\frac{e^{-\frac{x}{\lambda_{x}}}}{\lambda_{x}}
\left(1-e^{-\frac{x \sigma_{2}^{2}}{\sigma_{1}^{2} \lambda_{y}}}\right. 
\nonumber\\&\left. 
+\sqrt{\pi} \varphi_{6}(x) e^{\varphi_{6}(x)^{2}}
\left[\operatorname{erf}\left(\varphi_{6}(x)\right)-\operatorname{erf}\left(\varphi_{6}(x)+\sqrt{\frac{\sigma_{2}^{2}(\overline{\gamma}_{1}+\overline{\gamma}_{1}\overline{\gamma}_{2})}{\sigma_{1}^{2} \overline{\gamma}_{2} \lambda_{z}}}\right)\right] 
\right) 
dx,
\label{B13_I11}
\end{align}
where we define $\varphi_{6}(x) \triangleq \sqrt{\frac{\sigma_{2}^{2} \overline{\gamma}_{2} \lambda_{z}}{4 \sigma_{1}^{2}(\overline{\gamma}_{1}+\overline{\gamma}_{1}\overline{\gamma}_{2}) \lambda_{y}^{2}}} x$, and $\operatorname{erf}(\cdot)$ denotes the error function \cite[Eq. (8.250)]{TableOfIntegrals}. Furthermore, by using \cite[Eq. (3.322.1)]{TableOfIntegrals} again, we can arrive at
\begin{align}
I_{1,2}=&
\int_{\frac{\sigma_{1}^{2}(\overline{\gamma}_{1}+\overline{\gamma}_{1}\overline{\gamma}_{2}+\overline{\gamma}_{2})}{P_{\textrm{p}}}}^{\infty}
\frac{e^{-\frac{x}{\lambda_{x}}}}{\lambda_{x}}
\left(1-e^{-\frac{\sigma_{2}^{2} \overline{\gamma}_{2} / \lambda_{y}}{P_{\textrm{p}}-\frac{\sigma_{1}^{2}(\overline{\gamma}_{1}+\overline{\gamma}_{1}\overline{\gamma}_{2})}{x}}}\right. 
\nonumber\\&\left. 
+\sqrt{\pi} \varphi_{6}(x) e^{\varphi_{6}(x)^{2}}
\left[\operatorname{erf}\left(\varphi_{6}(x)\right)-\operatorname{erf}\left(\varphi_{6}(x)+\frac{\sqrt{\frac{\sigma_{1}^{2} \sigma_{2}^{2} \overline{\gamma}_{2}\left(\overline{\gamma}_{1}+\overline{\gamma}_{1} \overline{\gamma}_{2}\right)}{\lambda_{z}}}}{P_{\textrm{p}} x-\sigma_{1}^{2}\left(\overline{\gamma}_{1}+\overline{\gamma}_{1} \overline{\gamma}_{2}\right)}\right)\right] 
\right) 
dx.
\label{B14_I12}
\end{align}
Now we turn to the second term (i.e., $I_{2}$). After analyzing its integral region, we have
\begin{align}
I_{2}=&
\int_{0}^{\frac{\sigma_{2}^{2}(\overline{\gamma}_{1}+\overline{\gamma}_{1}\overline{\gamma}_{2})}{\sigma_{1}^{2} \overline{\gamma}_{2}}}
\int_{0}^{\varphi_{7}(z)}
\int_{0}^{x \sqrt{\frac{\sigma_{2}^{2} \overline{\gamma}_{2} z}{\sigma_{1}^{2}(\overline{\gamma}_{1}+\overline{\gamma}_{1}\overline{\gamma}_{2})}}}
\frac{e^{-\frac{y}{\lambda_{y}}}}{\lambda_{y}} d y \frac{e^{-\frac{x}{\lambda_{x}}}}{\lambda_{x}} d x
\frac{e^{-\frac{z}{\lambda_{z}}}}{\lambda_{z}} d z
\nonumber\\&
+\int_{0}^{\frac{\sigma_{2}^{2}(\overline{\gamma}_{1}+\overline{\gamma}_{1}\overline{\gamma}_{2})}{\sigma_{1}^{2} \overline{\gamma}_{2}}}
\int_{\varphi_{7}(z)}^{\varphi_{8}(z)}
\int_{0}^{\varphi_{8}(z)z-xz}
\frac{e^{-\frac{y}{\lambda_{y}}}}{\lambda_{y}} d y \frac{e^{-\frac{x}{\lambda_{x}}}}{\lambda_{x}} d x
\frac{e^{-\frac{z}{\lambda_{z}}}}{\lambda_{z}} d z
\triangleq I_{2,1}+I_{2,2},
\label{B15_I2}
\end{align}
where $\varphi_{7}(z)\triangleq \frac{\sigma_{1}^{2}(\overline{\gamma}_{1}+\overline{\gamma}_{1}\overline{\gamma}_{2})\left( 2 \sqrt{\frac{\sigma_{2}^{2} \overline{\gamma}_{2} z}{\sigma_{1}^{2}(\overline{\gamma}_{1}+\overline{\gamma}_{1}\overline{\gamma}_{2})}}+z\right) +\sigma_{2}^{2} \overline{\gamma}_{2}}{P_{\textrm{p}}\left( \sqrt{\frac{\sigma_{2}^{2} \overline{\gamma}_{2}z}{\sigma_{1}^{2}(\overline{\gamma}_{1}+\overline{\gamma}_{1}\overline{\gamma}_{2})}}+z\right) }$ and $\varphi_{8}(z)\triangleq \frac{\sigma_{1}^{2}(\overline{\gamma}_{1}+\overline{\gamma}_{1}\overline{\gamma}_{2})\left( 2 \sqrt{\frac{\sigma_{2}^{2} \overline{\gamma}_{2} z}{\sigma_{1}^{2}(\overline{\gamma}_{1}+\overline{\gamma}_{1}\overline{\gamma}_{2})}}+z\right) +\sigma_{2}^{2} \overline{\gamma}_{2}}{P_{\textrm{p}}z }$. After performing some algebraic arrangements, one can show that
\begin{align}
I_{2,1}=
\int_{0}^{\frac{\sigma_{2}^{2}(\overline{\gamma}_{1}+\overline{\gamma}_{1}\overline{\gamma}_{2})}{\sigma_{1}^{2} \overline{\gamma}_{2}}}
\left[
1-e^{-\frac{\varphi_{10}(z)}{\lambda_{x} P_{\textrm{p}}\left(\varphi_{9}(z)+z\right)}}
-\frac{1-e^{-\left(\frac{\varphi_{9}(z)}{\lambda_{y}}+\frac{1}{\lambda_{x}}\right) \frac{\varphi_{10}(z)}{P_{\textrm{p}}\left(\varphi_{9}(z)+z\right)}}}{\lambda_{x}\left(\frac{\varphi_{9}(z)}{\lambda_{y}}+\frac{1}{\lambda_{x}}\right)}
\right] 
\frac{e^{-\frac{z}{\lambda_{z}}}}{\lambda_{z}} dz,
\label{B16_I21}
\end{align}
\begin{align}
I_{2,2}=&
\int_{0}^{\frac{\sigma_{2}^{2}(\overline{\gamma}_{1}+\overline{\gamma}_{1}\overline{\gamma}_{2})}{\sigma_{1}^{2} \overline{\gamma}_{2}}}
\left[
e^{-\frac{\varphi_{10}(z)}{\lambda_{x} P_{\textrm{p}}\left(\varphi_{9}(z)+z\right)}}
-e^{-\frac{\varphi_{10}(z)}{\lambda_{x} P_{\textrm{p}} z}}
\right. \nonumber\\&\left.
-\frac{e^{-\frac{\varphi_{10}(z)}{\lambda_{y} P_{\textrm{p}}}}}{\lambda_{x}\left(\frac{z}{\lambda_{y}}-\frac{1}{\lambda_{x}}\right)}
\left(e^{\left(\frac{z}{\lambda_{y}}-\frac{1}{\lambda_{x}}\right) \frac{\varphi_{10}(z)}{P_{\textrm{p}} z}}-e^{\left(\frac{z}{\lambda_{y}}-\frac{1}{\lambda_{x}}\right) \frac{\varphi_{10}(z)}{P_{\textrm{p}}\left(\varphi_{9}(z)+z\right)}}\right) 
\right] 
\frac{e^{-\frac{z}{\lambda_{z}}}}{\lambda_{z}} dz,
\label{B17_I22}
\end{align}
where we define $\varphi_{9}(z)\triangleq \sqrt{\frac{\sigma_{2}^{2} \overline{\gamma}_{2} z}{\sigma_{1}^{2}(\overline{\gamma}_{1}+\overline{\gamma}_{1}\overline{\gamma}_{2})}}$ and $\varphi_{10}(z)\triangleq \sigma_{1}^{2}(\overline{\gamma}_{1}+\overline{\gamma}_{1} \overline{\gamma}_{2})\left(2 \varphi_{9}(z)+z\right)+\sigma_{2}^{2} \overline{\gamma}_{2}$. Finally, we turn to the last term (i.e., $I_{3}$). After analyzing its integral region, we can arrive at
\begin{align}
&I_{3}=
\int_{\frac{\sigma_{2}^{2}(\overline{\gamma}_{1}+\overline{\gamma}_{1}\overline{\gamma}_{2})}{\sigma_{1}^{2} \overline{\gamma}_{2}}}^{\infty}
\int_{0}^{\frac{\sigma_{1}^{2}(\overline{\gamma}_{1}+\overline{\gamma}_{1}\overline{\gamma}_{2}+\overline{\gamma}_{2})}{P_{\textrm{p}}}}
\int_{0}^{x \frac{\sigma_{2}^{2}}{\sigma_{1}^{2}}}
\frac{e^{-\frac{y}{\lambda_{y}}}}{\lambda_{y}} d y \frac{e^{-\frac{x}{\lambda_{x}}}}{\lambda_{x}} d x
\frac{e^{-\frac{z}{\lambda_{z}}}}{\lambda_{z}} d z
\nonumber\\&
+\int_{\frac{\sigma_{2}^{2}(\overline{\gamma}_{1}+\overline{\gamma}_{1}\overline{\gamma}_{2})}{\sigma_{1}^{2} \overline{\gamma}_{2}}}^{\infty}
\int_{\frac{\sigma_{1}^{2}(\overline{\gamma}_{1}+\overline{\gamma}_{1}\overline{\gamma}_{2}+\overline{\gamma}_{2})}{P_{\textrm{p}}}}^{\frac{(\overline{\gamma}_{1}+\overline{\gamma}_{1}\overline{\gamma}_{2}+\overline{\gamma}_{2})\left(\sigma_{2}^{2}+\sigma_{1}^{2} z\right)}{P_{\textrm{p}} z}}
\int_{0}^{\frac{(\overline{\gamma}_{1}+\overline{\gamma}_{1}\overline{\gamma}_{2}+\overline{\gamma}_{2})\left(\sigma_{2}^{2}+\sigma_{1}^{2} z\right)}{P_{\textrm{p}}}-xz}
\frac{e^{-\frac{y}{\lambda_{y}}}}{\lambda_{y}} d y \frac{e^{-\frac{x}{\lambda_{x}}}}{\lambda_{x}} d x
\frac{e^{-\frac{z}{\lambda_{z}}}}{\lambda_{z}} d z
\triangleq I_{3,1}+I_{3,2}.
\label{B18_I3}
\end{align}
By performing some algebraic manipulations, we have
\begin{align}
I_{3,1}=\left(1-\frac{1}{\left(\frac{\lambda_{x} \sigma_{2}^{2}}{\lambda_{y} \sigma_{1}^{2}}+1\right)}+\left[\frac{e^{-\frac{\sigma_{2}^{2}\left(\overline{\gamma}_{1}+\overline{\gamma}_{1} \overline{\gamma}_{2}+\overline{\gamma}_{2}\right)}{\lambda_{y} P_{\textrm{p}}}}}{\left(\frac{\lambda_{x} \sigma_{2}^{2}}{\lambda_{y} \sigma_{1}^{2}}+1\right)}-1\right] e^{-\frac{\sigma_{1}^{2}(\overline{\gamma}_{1}+\overline{\gamma}_{1}\overline{\gamma}_{2}+\overline{\gamma}_{2})}{\lambda_{x} P_{\textrm{p}}}}\right) e^{-\frac{\sigma_{2}^{2}(\overline{\gamma}_{1}+\overline{\gamma}_{1}\overline{\gamma}_{2})}{\lambda_{z} \sigma_{1} \overline{\gamma}_{2}}},
\label{B19_I31}
\end{align}
\begin{align}
&I_{3,2}=
\int_{\frac{\sigma_{2}^{2}(\overline{\gamma}_{1}+\overline{\gamma}_{1}\overline{\gamma}_{2})}{\sigma_{1}^{2} \overline{\gamma}_{2}}}^{\infty}
\left[
e^{-\frac{\sigma_{1}^{2}(\overline{\gamma}_{1}+\overline{\gamma}_{1}\overline{\gamma}_{2}+\overline{\gamma}_{2})}{\lambda_{x} P_{\textrm{p}}}}
-e^{-\frac{(\overline{\gamma}_{1}+\overline{\gamma}_{1}\overline{\gamma}_{2}+\overline{\gamma}_{2})\left(\sigma_{2}^{2}+\sigma_{1}^{2} z\right)}{\lambda_{x} P_{\textrm{p}} z}}
\right. \nonumber\\&\left. 
-\frac{e^{-\frac{(\overline{\gamma}_{1}+\overline{\gamma}_{1}\overline{\gamma}_{2}+\overline{\gamma}_{2})\left(\sigma_{2}^{2}+\sigma_{1}^{2} z\right)}{\lambda_{y} P_{\textrm{p}}}}}{\lambda_{x}\left(\frac{z}{\lambda_{y}}-\frac{1}{\lambda_{x}}\right)}\left(e^{\left(\frac{z}{\lambda_{y}}-\frac{1}{\lambda_{x}}\right) \frac{(\overline{\gamma}_{1}+\overline{\gamma}_{1}\overline{\gamma}_{2}+\overline{\gamma}_{2})\left(\sigma_{2}^{2}+\sigma_{1}^{2} z\right)}{P_{\textrm{p}} z}}-e^{\left(\frac{z}{\lambda_{y}}-\frac{1}{\lambda_{x}}\right) \frac{\sigma_{1}^{2}(\overline{\gamma}_{1}+\overline{\gamma}_{1}\overline{\gamma}_{2}+\overline{\gamma}_{2})}{P_{\textrm{p}}}}\right)
\right] 
\frac{e^{-\frac{z}{\lambda_{z}}}}{\lambda_{z}} d z.
\label{B20_I32}
\end{align}
Combining the foregoing results, we complete the derivation of the SOP for the BC-NOMA scheme. More concise expressions cannot be achieved due to the complicated integrals.

\section{Derivation of the DMT Performance}
\setcounter{equation}{0}
For each fading block, if $|h_{\textrm{A}}|^2/\sigma_{\textrm{A}}^2\ge |h_{\textrm{B}}|^2/\sigma_{\textrm{B}}^2$, we define $r_1\triangleq r_{\textrm{A}}$ and $r_2\triangleq r_{\textrm{B}}$. On the contrary, if $|h_{\textrm{A}}|^2/\sigma_{\textrm{A}}^2< |h_{\textrm{B}}|^2/\sigma_{\textrm{B}}^2$, we define $r_1\triangleq r_{\textrm{B}}$ and $r_2\triangleq r_{\textrm{A}}$. The DMT performance can be obtained by substituting the multiplexing gains (i.e., $r_1$ and $r_2$) for the threshold SNRs (i.e., $\overline{\gamma}_{1}$ and $\overline{\gamma}_{2}$) in the expression of the SOP (i.e., $P_{\textrm{out}}$) in the high SNR region, and then by examining the decaying rate of $P_{\textrm{out}}$ with an increase of the maximum allowed total power (i.e., $P_{\textrm{p}}$). The substitutions are conducted by using the following two equations.
\begin{align}
\overline{\gamma}_{1}=2^{R_{1}}-1=\left(1+\frac{P_{\textrm{p}} \lambda_{x}}{\sigma_{1}^{2}}\right)^{r_{1}}-1 \rightarrow\left(\frac{\lambda_{x}}{\sigma_{1}^{2}}\right)^{r_{1}} P_{\textrm{p}}^{r_{1}},
\label{D1_gamma1}
\end{align}
\begin{align}
\overline{\gamma}_{2}=2^{R_{2}}-1=\left(1+\frac{P_{\textrm{p}} \lambda_{y}}{\sigma_{2}^{2}}\right)^{r_{2}}-1 \rightarrow\left(\frac{\lambda_{y}}{\sigma_{2}^{2}}\right)^{r_{2}} P_{\textrm{p}}^{r_{2}},
\label{D2_gamma2}
\end{align}
which are developed from \eqref{E21_MultiplexingGainA} and \eqref{E22_MultiplexingGainB}.

In what follows, we determine the DMT performance of the NC-NOMA scheme, the IR-NOMA scheme, and the BC-NOMA scheme, respectively. The DMT performance of the CR-NOMA scheme can be obtained by directly replacing $r_\textrm{A}$ and $r_\textrm{B}$ with $2r_\textrm{A}$ and $2r_\textrm{B}$, respectively, in the derived DMT performance of the IR-NOMA scheme. This is because the SOP of the CR-NOMA scheme is the same as that of the IR-NOMA scheme, whereas the CR-NOMA scheme involves a cooperative transmission phase after each direct transmission phase, which halves the data rates. Note that the IR-NOMA scheme also involves additional time slots. However, according to \cite[Claim 3]{Laneman04TIT}, there is no rate loss in the high SNR region for an IR scheme. This is because in the high SNR region, the direct transmission phase is sufficient for avoiding information outage in most fading blocks, and thus the cooperative transmission phase is rarely activated for the IR scheme.

\noindent \textbf{C-1: The DMT performance of the NC-NOMA Scheme}

It follows from \eqref{B3_Pout'NC1} that as $P_{\textrm{p}} \rightarrow \infty$, we have $P_{\textrm{out,NC1}}^{\prime} \rightarrow \frac{\sigma_{2}^{2} \overline{\gamma}_{2}}{\lambda_{y} P_{\textrm{p}}}$. Furthermore, by making use of \eqref{D2_gamma2}, it is ready to determine that $P_{\textrm{out,NC1}}^{\prime} \rightarrow \left(\frac{\lambda_{y}}{\sigma_{2}^{2}}\right)^{r_{2}-1} \frac{1}{P_{\textrm{p}}^{1-r_{2}}}$. Now we turn to $P_{\textrm{out,NC2}}^{\prime}$. As $P_{\textrm{p}} \rightarrow \infty$, by noting that both the upper limit and the lower limit of the integral in \eqref{B4_Pout'NC2} approach to zero, we have
\begin{align}
P_{\textrm{out,NC2}}^{\prime} \rightarrow \frac{1}{\lambda_{y}} \int_{\frac{\sigma_{2}^{2} \overline{\gamma}_{2}}{P_{\textrm{p}}}}^{\frac{\sigma_{2}^{2}(\overline{\gamma}_{1}+\overline{\gamma}_{1} \overline{\gamma}_{2}+\overline{\gamma}_{2}) }{P_{\textrm{p}}}} 1-e^{-\frac{y \sigma_{1}^{2} (\overline{\gamma}_{1}+\overline{\gamma}_{1} \overline{\gamma}_{2})}{\lambda_{x}\left(y P_{\textrm{p}}-\sigma_{2}^{2} \overline{\gamma}_{2}\right)}} d y \triangleq P_{\textrm{out,NC2}}^{\prime \prime}.
\label{D3_PoutNC2'}
\end{align}
By applying the change of variables $y^{\prime}=y P_{\textrm{p}}-\sigma_{2}^{2} \overline{\gamma}_{2}$ and using $\int e^{-\frac{a}{x}} d x=a \Gamma\left(-1, \frac{a}{x}\right)$, where $\Gamma(\cdot,\cdot)$ denotes the incomplete Gamma function \cite[Eq. (6.5.3)]{HandbookOf}, one can further show that
\begin{align}
P_{\textrm{out,NC2}}^{\prime \prime}=\frac{\sigma_{2}^{2}(\overline{\gamma}_{1}+\overline{\gamma}_{1} \overline{\gamma}_{2})}{\lambda_{y} P_{\textrm{p}}}\left(1-e^{-\frac{\sigma_{1}^{2} (\overline{\gamma}_{1}+\overline{\gamma}_{1} \overline{\gamma}_{2})}{\lambda_{x} P_{\textrm{p}}}} \frac{\sigma_{1}^{2} \overline{\gamma}_{2}}{\lambda_{x} P_{\textrm{p}}} \Gamma\left(-1, \frac{\sigma_{1}^{2} \overline{\gamma}_{2}}{\lambda_{x} P_{\textrm{p}}}\right)\right).
\label{D4_PoutNC2''}
\end{align}
Next, by making use of \cite[Eq. (6.5.19)]{HandbookOf} and \cite[Eq. (5.1.11)]{HandbookOf}, we can arrive at
\begin{align}
P_{\textrm{out,NC2}}^{\prime \prime}&\rightarrow \frac{\sigma_{2}^{2}(\overline{\gamma}_{1}+\overline{\gamma}_{1} \overline{\gamma}_{2})}{\lambda_{y} P_{\textrm{p}}}
\left(1-e^{-\frac{\sigma_{1}^{2}(\overline{\gamma}_{1}+\overline{\gamma}_{1}\overline{\gamma}_{2}+\overline{\gamma}_{2})}{\lambda_{x} P_{\textrm{p}}}}+e^{-\frac{\sigma_{1}^{2} (\overline{\gamma}_{1}+\overline{\gamma}_{1} \overline{\gamma}_{2})}{\lambda_{x} P_{\textrm{p}}}} \frac{\sigma_{1}^{2} \overline{\gamma}_{2}}{\lambda_{x} P_{\textrm{p}}} \ln \left(\frac{\lambda_{x} P_{\textrm{p}}}{\sigma_{1}^{2} \overline{\gamma}_{2}}\right)\right)
\nonumber\\&
\rightarrow
\frac{\sigma_{1}^{2} \sigma_{2}^{2}(\overline{\gamma}_{1}+\overline{\gamma}_{1} \overline{\gamma}_{2})\left(\overline{\gamma}_{1}+\overline{\gamma}_{1} \overline{\gamma}_{2}+\overline{\gamma}_{2}\right)}{\lambda_{x} \lambda_{y} P_{\textrm{p}}^{2}} \rightarrow\left(\frac{\lambda_{x}}{\sigma_{1}^{2}}\right)^{2 r_{1}-1}\left(\frac{\lambda_{y}}{\sigma_{2}^{2}}\right)^{2 r_{2}-1} \frac{1}{P_{\textrm{p}}^{2-2 r_{1}-2 r_{2}}}.
\label{D5_PoutNC2''2}
\end{align}
The last step of \eqref{D5_PoutNC2''2} is obtained by using \eqref{D1_gamma1} and \eqref{D2_gamma2}. Finally, note that the decaying rate of $P_{\textrm{out,NC}}^{\prime}$ is determined by the dominating term. By combining $P_{\textrm{out,NC1}}^{\prime} \rightarrow \left(\frac{\lambda_{y}}{\sigma_{2}^{2}}\right)^{r_{2}-1} \frac{1}{P_{\textrm{p}}^{1-r_{2}}}$ with \eqref{D5_PoutNC2''2} and noting that both user A and user B can act as user 2, we can conclude that the DMT performance of the NC-NOMA scheme is $\min\left\lbrace1-r_{\textrm{A}},1-r_{\textrm{B}},2-2r_{\textrm{A}}-2r_{\textrm{B}}\right\rbrace $.

\noindent \textbf{C-2: The DMT performance of the IR-NOMA Scheme}

By noting that $\left(\frac{1}{y}+\frac{\sigma_{1}^{2} \overline{\gamma}_{1}}{\sigma_{2}^{2} x}\right)^{-1} \leq y$, it follows from \eqref{B6_Q11_1} that
\begin{align}
Q_{1,1} \leq \int_{0}^{\frac{\sigma_{2}^{2} \overline{\gamma}_{2}}{P_{\textrm{p}}}} \int_{\frac{\sigma_{1}^{2}}{\sigma_{2}^{2}}y}^{\infty} \int_{0}^{y} \frac{e^{-\frac{z}{\lambda_{z}}}}{\lambda_{z}} d z \frac{e^{-\frac{x}{\lambda_{x}}}}{\lambda_{x}} d x \frac{e^{-\frac{y}{\lambda_{y}}}}{\lambda_{y}} d y \triangleq Q_{1,1}^{\prime}.
\label{D6_Q11}
\end{align}
Next, by noting that the upper limits of variables $y$ and $z$ in \eqref{D6_Q11} approach to zero as $P_{\textrm{p}} \rightarrow \infty$ and then by using \eqref{D1_gamma1} as well as \eqref{D2_gamma2}, we have
\begin{align}
Q_{1,1}^{\prime} \rightarrow \int_{0}^{\frac{\sigma_{2}^{2} \overline{\gamma}_{2}}{P_{\textrm{p}}}} \int_{0}^{\infty} \int_{0}^{y} \frac{1}{\lambda_{z}} d z \frac{e^{-\frac{x}{\lambda_{x}}}}{\lambda_{x}} d x \frac{1}{\lambda_{y}} d y
=\frac{\left(\sigma_{2}^{2}\right)^{2}}{2 \lambda_{y} \lambda_{z}} \frac{\left(\overline{\gamma}_{2}\right)^{2}}{P_{\textrm{p}}^{2}} \rightarrow \frac{\left(\sigma_{2}^{2}\right)^{2}}{2 \lambda_{y} \lambda_{z}}\left(\frac{\lambda_{y}}{\sigma_{2}^{2}}\right)^{2 r_{2}} \frac{1}{P_{\textrm{p}}^{2-2 r_{2}}}.
\label{D7_Q11'}
\end{align}
In a similar way, we can derive from \eqref{B8_Q12} that
\begin{align}
Q_{1,2} &\leq \int_{\frac{\sigma_{2}^{2} \overline{\gamma}_{2}}{P_{\textrm{p}}}}^{\frac{\sigma_{2}^{2}(\overline{\gamma}_{1}+\overline{\gamma}_{1} \overline{\gamma}_{2}+\overline{\gamma}_{2})}{P_{\textrm{p}}}} \int_{\frac{\sigma_{1}^{2}}{\sigma_{2}^{2}}y}^{\frac{\sigma_{1}^{2} (\overline{\gamma}_{1}+\overline{\gamma}_{1} \overline{\gamma}_{2})}{P_{\textrm{p}}-\frac{\sigma_{2}^{2} \overline{\gamma}_{2}}{y}}} \int_{0}^{y} \frac{e^{-\frac{z}{\lambda_{z}}}}{\lambda_{z}} d z \frac{e^{-\frac{x}{\lambda_{x}}}}{\lambda_{x}} d x \frac{e^{-\frac{y}{\lambda_{y}}}}{\lambda_{y}} d y
\nonumber\\&
\rightarrow \int_{\frac{\sigma_{2}^{2} \overline{\gamma}_{2}}{P_{\textrm{p}}}}^{\frac{\sigma_{2}^{2}(\overline{\gamma}_{1}+\overline{\gamma}_{1} \overline{\gamma}_{2}+\overline{\gamma}_{2})}{P_{\textrm{p}}}} \int_{\frac{\sigma_{1}^{2}}{\sigma_{2}^{2}}y}^{\frac{\sigma_{1}^{2} (\overline{\gamma}_{1}+\overline{\gamma}_{1} \overline{\gamma}_{2})}{P_{\textrm{p}}-\frac{\sigma_{2}^{2} \overline{\gamma}_{2}}{y}}} \int_{0}^{y}
\frac{1}{\lambda_{z}} d z \frac{e^{-\frac{x}{\lambda_{x}}}}{\lambda_{x}} d x \frac{1}{\lambda_{y}} d y
\nonumber\\&
=\frac{1}{\lambda_{y}\lambda_{z}}\int_{\frac{\sigma_{2}^{2} \overline{\gamma}_{2}}{P_{\textrm{p}}}}^{\frac{\sigma_{2}^{2}(\overline{\gamma}_{1}+\overline{\gamma}_{1} \overline{\gamma}_{2}+\overline{\gamma}_{2})}{P_{\textrm{p}}}}
\left( 1-e^{-\frac{\sigma_{1}^{2} (\overline{\gamma}_{1}+\overline{\gamma}_{1} \overline{\gamma}_{2}) / \lambda_{x}}{P_{\textrm{p}}-\frac{\sigma_{2}^{2} \overline{\gamma}_{2}}{y}}}\right) y
d y\triangleq Q_{1,2}^{\prime}.
\label{D8_Q12}
\end{align}
Furthermore, one can readily show that
\begin{align}
Q_{1,2}^{\prime} &\leq
\frac{1}{\lambda_{y}\lambda_{z}}\int_{\frac{\sigma_{2}^{2} \overline{\gamma}_{2}}{P_{\textrm{p}}}}^{\frac{\sigma_{2}^{2}(\overline{\gamma}_{1}+\overline{\gamma}_{1} \overline{\gamma}_{2}+\overline{\gamma}_{2})}{P_{\textrm{p}}}}
y dy=\frac{\left(\sigma_{2}^{2}\right)^{2}\left((\overline{\gamma}_{1}+\overline{\gamma}_{1} \overline{\gamma}_{2})^{2}+2 \overline{\gamma}_{2}(\overline{\gamma}_{1}+\overline{\gamma}_{1} \overline{\gamma}_{2})\right)}{2 \lambda_{y} \lambda_{z} P_{\textrm{p}}^{2}}
\nonumber\\&
\rightarrow
\frac{\left(\sigma_{2}^{2}\right)^{2}}{2 \lambda_{y} \lambda_{z}}\left(\frac{\lambda_{x}}{\sigma_{1}^{2}}\right)^{2 r_{1}}\left(\frac{\lambda_{y}}{\sigma_{2}^{2}}\right)^{2 r_{2}} \frac{1}{P_{\textrm{p}}^{2-2 r_{1}-2 r_{2}}}.
\label{D9_Q12'}
\end{align}
The last step of \eqref{D9_Q12'} is obtained by using \eqref{D1_gamma1} and \eqref{D2_gamma2} as before. Now we turn to $Q_{2,1}$. By noting that the upper limits of variables $x$ and $y$ in \eqref{B9_Q21} approach to zero as $P_{\textrm{p}} \rightarrow \infty$, we have
\begin{align}
Q_{2,1} &\rightarrow \int_{0}^{\frac{\sigma_{2}^{2}(\overline{\gamma}_{1}+\overline{\gamma}_{1}\overline{\gamma}_{2}+\overline{\gamma}_{2})}{P_{\textrm{p}}}} \int_{0}^{\frac{\sigma_{1}^{2}(\overline{\gamma}_{1}+\overline{\gamma}_{1}\overline{\gamma}_{2}+\overline{\gamma}_{2})}{P_{\textrm{p}}}} \int_{0}^{\infty} \frac{e^{-\frac{z}{\lambda_{z}}}}{\lambda_{z}} d z \frac{1}{\lambda_{x}} d x \frac{1}{\lambda_{y}} d y
\nonumber\\&=
\frac{\sigma_{1}^{2} \sigma_{2}^{2}(\overline{\gamma}_{1}+\overline{\gamma}_{1}\overline{\gamma}_{2}+\overline{\gamma}_{2})^{2}}{\lambda_{x} \lambda_{y} P_{\textrm{p}}^{2}} \rightarrow\left(\frac{\lambda_{x}}{\sigma_{1}^{2}}\right)^{2 r_{1}-1}\left(\frac{\lambda_{y}}{\sigma_{2}^{2}}\right)^{2 r_{2}-1} \frac{1}{P_{\textrm{p}}^{2-2 r_{1}-2 r_{2}}}.
\label{D10_Q21}
\end{align}
Finally, we turn to $Q_{2,2}$. Note that as $P_{\textrm{p}} \rightarrow \infty$, the upper limit of variable $y$ and the lower limit of variable $z$ in \eqref{B10_Q22} approach to zero. By applying the change of variables $x=x^{\prime} \frac{\sigma_{1}^{2}(\overline{\gamma}_{1}+\overline{\gamma}_{1}\overline{\gamma}_{2}+\overline{\gamma}_{2})}{P_\textrm{p}}$ and $y=y^{\prime} \frac{\sigma_{2}^{2} \overline{\gamma}_{2}}{P_{\textrm{p}}\left(1-\frac{\overline{\gamma}_{1}+\overline{\gamma}_{1}\overline{\gamma}_{2}}{x(\overline{\gamma}_{1}+\overline{\gamma}_{1}\overline{\gamma}_{2}+\overline{\gamma}_{2})}\right)}$, one can show that
\begin{align}
Q_{2,2} \rightarrow&\frac{\sigma_{1}^{2} \sigma_{2}^{2}(\overline{\gamma}_{1}+\overline{\gamma}_{1}\overline{\gamma}_{2}+\overline{\gamma}_{2})}{\lambda_{x} \lambda_{y} P_{\textrm{p}}^{2}} \int_{1}^{\infty} \int_{0}^{1} \int_{0}^{\frac{\overline{\gamma}_{2} \sigma_{2}^{2}\left(1-\frac{\overline{\gamma}_{2} y}{x(\overline{\gamma}_{1}+\overline{\gamma}_{1}\overline{\gamma}_{2}+\overline{\gamma}_{2})-(\overline{\gamma}_{1}+\overline{\gamma}_{1}\overline{\gamma}_{2})}\right)}{P_{\textrm{p}}\left(1-\frac{1}{x}\right)\left(1+\frac{\overline{\gamma}_{1}\overline{\gamma}_{2} y}{x(\overline{\gamma}_{1}+\overline{\gamma}_{1}\overline{\gamma}_{2}+\overline{\gamma}_{2})-(\overline{\gamma}_{1}+\overline{\gamma}_{1}\overline{\gamma}_{2})}\right)}} \frac{e^{-\frac{z}{\lambda_{z}}}}{\lambda_{z}} d z d y
\nonumber\\&\times \frac{\overline{\gamma}_{2}}{1-\frac{\overline{\gamma}_{1}+\overline{\gamma}_{1} \overline{\gamma}_{2}}{x(\overline{\gamma}_{1}+\overline{\gamma}_{1} \overline{\gamma}_{2}+\overline{\gamma}_{2})}} e^{-x \frac{\sigma_{1}^{2}(\overline{\gamma}_{1}+\overline{\gamma}_{1}\overline{\gamma}_{2}+\overline{\gamma}_{2})}{\lambda_{x} P_{\textrm{p}}}}d x
\nonumber\\\le&
\frac{\sigma_{1}^{2} \sigma_{2}^{2}\left( \overline{\gamma}_{1}+\overline{\gamma}_{1}\overline{\gamma}_{2}+\overline{\gamma}_{2}\right)^2 }{\lambda_{x} \lambda_{y} P_{\textrm{p}}^{2}} \int_{1}^{\infty} \int_{0}^{1} \int_{0}^{\frac{\overline{\gamma}_{2} \sigma_{2}^{2}}{P_{\textrm{p}}\left(1-\frac{1}{x}\right)}} \frac{e^{-\frac{z}{\lambda_{z}}}}{\lambda_{z}} d z d y
e^{-x \frac{\sigma_{1}^{2}\left( \overline{\gamma}_{1}+\overline{\gamma}_{1}\overline{\gamma}_{2}+\overline{\gamma}_{2}\right)}{\lambda_{x} P_{\textrm{p}}}}d x
\nonumber\\=&
\frac{\sigma_{1}^{2} \sigma_{2}^{2}\left( \overline{\gamma}_{1}+\overline{\gamma}_{1}\overline{\gamma}_{2}+\overline{\gamma}_{2}\right)^2 }{\lambda_{x} \lambda_{y} P_{\textrm{p}}^{2}}\int_{0}^{\infty}\left(1-e^{-\frac{ \overline{\gamma}_{2} \sigma_{2}^{2}}{\lambda_{z} P_{\textrm{p}}\left(1-\frac{1}{x+1}\right)}}\right) e^{-(x+1)\frac{\sigma_{1}^{2}\left( \overline{\gamma}_{1}+\overline{\gamma}_{1}\overline{\gamma}_{2}+\overline{\gamma}_{2}\right)}{\lambda_{x} P_{\textrm{p}}}}dx\triangleq Q_{2,2}^{\prime}.
\label{D11_Q22}
\end{align}
Next, as $P_{\textrm{p}} \rightarrow \infty$, we can arrive at
\begin{align}
Q_{2,2}^{\prime} &\rightarrow
\frac{\sigma_{1}^{2} \sigma_{2}^{2}\left( \overline{\gamma}_{1}+\overline{\gamma}_{1}\overline{\gamma}_{2}+\overline{\gamma}_{2}\right)^2 }{\lambda_{x} \lambda_{y} P_{\textrm{p}}^{2}}
\int_{0}^{\infty}\left(1-e^{-\frac{ \overline{\gamma}_{2} \sigma_{2}^{2}}{\lambda_{z} P_{\textrm{p}}x}}\right) e^{-x\frac{\sigma_{1}^{2}\left( \overline{\gamma}_{1}+\overline{\gamma}_{1}\overline{\gamma}_{2}+\overline{\gamma}_{2}\right)}{\lambda_{x} P_{\textrm{p}}}}dx
\nonumber\\&=
\frac{\sigma_{1}^{2} \sigma_{2}^{2}\left( \overline{\gamma}_{1}+\overline{\gamma}_{1}\overline{\gamma}_{2}+\overline{\gamma}_{2}\right)^2 }{\lambda_{x} \lambda_{y} P_{\textrm{p}}^{2}}
 \underbrace{\left(\frac{\lambda_{x} P_{\textrm{p}}}{\sigma_{1}^{2}(\overline{\gamma}_{1}+\overline{\gamma}_{1} \overline{\gamma}_{2}+\overline{\gamma}_{2})}-\int_{0}^{\infty} e^{-\frac{\overline{\gamma}_{2} \sigma_{2}^{2}}{\lambda_{z} P_{\textrm{p}} x}-x \frac{\sigma_{1}^{2}(\overline{\gamma}_{1}+\overline{\gamma}_{1}\overline{\gamma}_{2}+\overline{\gamma}_{2})}{\lambda_{x} P_{\textrm{p}}}} d x\right)}_{\xi} .
\label{D12_Q22'}
\end{align}
By making use of \cite[Eq. (3.324)]{TableOfIntegrals} and \cite[Eq. (9.6.9)]{HandbookOf}, as $P_{\textrm{p}} \rightarrow \infty$, we have $\xi\rightarrow0$. Therefore, it follows from \eqref{D12_Q22'} that 
\begin{align}
Q_{2,2}^{\prime} &\le
\frac{\sigma_{1}^{2} \sigma_{2}^{2}\left( \overline{\gamma}_{1}+\overline{\gamma}_{1}\overline{\gamma}_{2}+\overline{\gamma}_{2}\right)^2 }{\lambda_{x} \lambda_{y} P_{\textrm{p}}^{2}}
\rightarrow
\left(\frac{\lambda_{x}}{\sigma_{1}^{2}}\right)^{2 r_{1}-1}\left(\frac{\lambda_{y}}{\sigma_{2}^{2}}\right)^{2 r_{2}-1} \frac{1}{P_{\textrm{p}}^{2-2 r_{1}-2 r_{2}}}.
\label{D13_Q22'2}
\end{align}
The last step of \eqref{D13_Q22'2} is obtained by using \eqref{D1_gamma1} and \eqref{D2_gamma2}. Note that the decaying rate of $P_{\textrm{out,CR/IR}}^{\prime}$ is determined by the dominating term. According to the foregoing results, the decaying rates of $Q_{1,1}$, $Q_{1,2}$, and $Q_{2,2}$ are faster than or at least the same as that of $Q_{2,1}$, which decays proportionally to $\frac{1}{P_{\textrm{p}}^{2-2 r_{1}-2 r_{2}}}$. Therefore, the DMT performance of the IR-NOMA scheme is $(2-2r_{\textrm{A}}-2r_{\textrm{B}})$, whereas the DMT performance of the CR-NOMA scheme is $(2-4r_{\textrm{A}}-4r_{\textrm{B}})$.

\noindent \textbf{C-3: The DMT performance of the BC-NOMA Scheme}

By noting that the upper limits of variables $x$ and $y$ in $I_{1,1}$ in \eqref{B12_I1} approach to zero as $P_{\textrm{p}} \rightarrow \infty$, one can show that 
\begin{align}
I_{1,1} \rightarrow \frac{1}{\lambda_{x} \lambda_{y}} \int_{0}^{\frac{\sigma_{1}^{2}(\overline{\gamma}_{1}+\overline{\gamma}_{1}\overline{\gamma}_{2}+\overline{\gamma}_{2})}{P_{\textrm{p}}}} \int_{0}^{\frac{x \sigma_{2}^{2}}{\sigma_{1}^{2}}}\left(1-e^{-\left(\frac{y}{x}\right)^{2} \frac{\sigma_{1}^{2}(\overline{\gamma}_{1}+\overline{\gamma}_{1}\overline{\gamma}_{2})}{\lambda_{z} \sigma_{2}^{2} \overline{\gamma}_{2}}}\right) d y d x \triangleq I_{1,1}^{\prime}.
\label{D14_I11}
\end{align}
Next, by applying the change of variables $y=y^{\prime} \frac{x \sigma_{2}^{2}}{\sigma_{1}^{2}}$ and using \cite[Eq. (3.321.2)]{TableOfIntegrals}, we have
\begin{align}
I_{1,1}^{\prime}&=
\left( 1-\sqrt{\frac{\pi \lambda_{z} \sigma_{1}^{2} \overline{\gamma}_{2}}{4 \sigma_{2}^{2}\left(\overline{\gamma}_{1}+\overline{\gamma}_{1} \overline{\gamma}_{2}\right)}} \operatorname{erf}\left( \sqrt{\frac{\sigma_{2}^{2}\left(\overline{\gamma}_{1}+\overline{\gamma}_{1} \overline{\gamma}_{2}\right)}{\lambda_{z} \sigma_{1}^{2} \overline{\gamma}_{2}}}\right) \right) \frac{\sigma_{1}^{2} \sigma_{2}^{2}\left(\overline{\gamma}_{1}+\overline{\gamma}_{1} \overline{\gamma}_{2}+\overline{\gamma}_{2}\right)^{2}}{2 \lambda_{x} \lambda_{y} P_{\textrm{p}}^{2}}
\nonumber\\&\rightarrow
\frac{1}{2}\left(\frac{\lambda_{x}}{\sigma_{1}^{2}}\right)^{2 r_{1}-1}\left(\frac{\lambda_{y}}{\sigma_{2}^{2}}\right)^{2 r_{2}-1} \frac{1}{P_{\textrm{p}}^{2-2 r_{1}-2 r_{2}}}.
\label{D15_I11'}
\end{align}
The last step of \eqref{D15_I11'} is obtained by using \eqref{D1_gamma1}, \eqref{D2_gamma2}, and $\operatorname{erf}(\infty)=1$. Now we turn to $I_{1,2}$ in \eqref{B12_I1}. By removing the exponential terms in $I_{1,2}$, one can show that 
\begin{align}
I_{1,2} &\leq \frac{1}{\lambda_{x} \lambda_{y} \lambda_{z}} \int_{\frac{\sigma_{1}^{2}\left( \overline{\gamma}_{1}+\overline{\gamma}_{1}\overline{\gamma}_{2}+\overline{\gamma}_{2}\right) }{P_{\textrm{p}}}}^{\infty}
\int_{0}^{\frac{\sigma_{2}^{2} \overline{\gamma}_{2}}{P_{\textrm{p}}-\frac{\sigma_{1}^{2}(\overline{\gamma}_{1}+\overline{\gamma}_{1}\overline{\gamma}_{2})}{x}}}
\int_{0}^{\left(\frac{y}{x}\right)^{2} \frac{\sigma_{1}^{2}(\overline{\gamma}_{1}+\overline{\gamma}_{1}\overline{\gamma}_{2})}{\sigma_{2}^{2} \overline{\gamma}_{2}}} d z d y d x
\nonumber\\&
=\frac{\left(\sigma_{2}^{2}\right)^{2}\left(\overline{\gamma}_{1}+\overline{\gamma}_{1} \overline{\gamma}_{2}\right)(\overline{\gamma}_{1}+\overline{\gamma}_{1} \overline{\gamma}_{2}+2 \overline{\gamma}_{2})}{6 \lambda_{x} \lambda_{y} \lambda_{z} P_{\textrm{p}}^{2}} \rightarrow \frac{\left(\sigma_{2}^{2}\right)^{2}}{6 \lambda_{x} \lambda_{y} \lambda_{z}}\left(\frac{\lambda_{x}}{\sigma_{1}^{2}}\right)^{2 r_{1}}\left(\frac{\lambda_{y}}{\sigma_{2}^{2}}\right)^{2 r_{2}} \frac{1}{P_{\textrm{p}}^{2-2 r_{1}-2 r_{2}}}.
\label{D16_I12}
\end{align}
The last step of \eqref{D16_I12} is obtained by using \eqref{D1_gamma1} and \eqref{D2_gamma2}. Next, we consider $I_{2,1}$. It is ready to derive from \eqref{B15_I2} that
\begin{align}
I_{2,1} &\leq \int_{0}^{\frac{\sigma_{2}^{2}(\overline{\gamma}_{1}+\overline{\gamma}_{1}\overline{\gamma}_{2})}{\sigma_{1}^{2}\overline{\gamma}_{2}}} \int_{0}^{\varphi_{7}(z)} \int_{0}^{x \sqrt{\frac{\sigma_{2}^{2} \overline{\gamma}_{2} z}{\sigma_{1}^{2}(\overline{\gamma}_{1}+\overline{\gamma}_{1}\overline{\gamma}_{2})}}} \frac{1}{\lambda_{y}} d y \frac{1}{\lambda_{x}} d x \frac{e^{-\frac{z}{\lambda_{z}}}}{\lambda_{z}} d z
\nonumber\\&=
\frac{1}{2 \lambda_{x} \lambda_{y}} \int_{0}^{\frac{\sigma_{2}^{2}(\overline{\gamma}_{1}+\overline{\gamma}_{1}\overline{\gamma}_{2})}{\sigma_{1}^{2} \overline{\gamma}_{2}}}\left(\varphi_{7}(z)\right)^{2} \sqrt{\frac{\sigma_{2}^{2} \overline{\gamma}_{2} z}{\sigma_{1}^{2}(\overline{\gamma}_{1}+\overline{\gamma}_{1} \overline{\gamma}_{2})}} \frac{e^{-\frac{z}{\lambda_{z}}}}{\lambda_{z}} d z \triangleq I_{2,1}^{\prime}.
\label{D17_I21}
\end{align}
Furthermore, applying the change of variables $z=z^{\prime} \frac{\sigma_{2}^{2}(\overline{\gamma}_{1}+\overline{\gamma}_{1}\overline{\gamma}_{2})}{\sigma_{1}^{2} \overline{\gamma}_{2}}$, we can arrive at
\begin{align}
I_{2,1}^{\prime}=
\frac{\left(\sigma_{2}^{2}\right)^{2}(\overline{\gamma}_{1}+\overline{\gamma}_{1} \overline{\gamma}_{2})}{2 \lambda_{x} \lambda_{y} \lambda_{z} \overline{\gamma}_{2} P_{\textrm{p}}^{2}}
\int_{0}^{1}
\underbrace{\frac{\left(\left(\overline{\gamma}_{1}+\overline{\gamma}_{1} \overline{\gamma}_{2}\right)\left(2 \sqrt{z}+z \frac{\left(\overline{\gamma}_{1}+\overline{\gamma}_{1} \overline{\gamma}_{2}\right)}{\overline{\gamma}_{2}}\right)+\overline{\gamma}_{2}\right)^{2}}{\left(\sqrt{z}+z \frac{\left(\overline{\gamma}_{1}+\overline{\gamma}_{1} \overline{\gamma}_{2}\right)}{\overline{\gamma}_{2}}\right)^{2}} \sqrt{z}}_{\chi(z)}
e^{-z \frac{\sigma_{2}^{2}(\overline{\gamma}_{1}+\overline{\gamma}_{1}\overline{\gamma}_{2})}{\lambda_{z} \sigma_{1}^{2} \overline{\gamma}_{2}}} d z.
\label{D18_I21'}
\end{align}
Note that
\begin{align}
\chi(z) &\leq \frac{\left(\left(\overline{\gamma}_{1}+\overline{\gamma}_{1} \overline{\gamma}_{2}\right)\left(2 \sqrt{z}+2 z \frac{\left(\overline{\gamma}_{1}+\overline{\gamma}_{1} \overline{\gamma}_{2}\right)}{\overline{\gamma}_{2}}\right)+\overline{\gamma}_{2}\right)^{2}}{\left(\sqrt{z}+z \frac{\left(\overline{\gamma}_{1}+\overline{\gamma}_{1} \overline{\gamma}_{2}\right)}{\overline{\gamma}_{2}}\right)^{2}} \sqrt{z}
\nonumber\\&=\left(2\left(\overline{\gamma}_{1}+\overline{\gamma}_{1} \overline{\gamma}_{2}\right)+\frac{\overline{\gamma}_{2}}{\sqrt{z}+z \frac{\left(\overline{\gamma}_{1}+\overline{\gamma}_{1} \overline{\gamma}_{2}\right)}{\overline{\gamma}_{2}}}\right)^{2} \sqrt{z} 
\nonumber\\&\leq\left(2\left(\overline{\gamma}_{1}+\overline{\gamma}_{1} \overline{\gamma}_{2}\right)+\frac{\overline{\gamma}_{2}}{\sqrt{z}}\right)^{2} \sqrt{z}
=\left(2\left(\overline{\gamma}_{1}+\overline{\gamma}_{1} \overline{\gamma}_{2}\right)+\frac{\overline{\gamma}_{2}}{\sqrt{z}}\right)\left(2\left(\overline{\gamma}_{1}+\overline{\gamma}_{1} \overline{\gamma}_{2}\right) \sqrt{z}+\overline{\gamma}_{2}\right)
\nonumber\\&\leq
\left(2\left(\overline{\gamma}_{1}+\overline{\gamma}_{1} \overline{\gamma}_{2}\right)+\frac{\overline{\gamma}_{2}}{\sqrt{z}}\right)\left(2\left(\overline{\gamma}_{1}+\overline{\gamma}_{1} \overline{\gamma}_{2}\right)+\overline{\gamma}_{2}\right).
\label{D19_chi}
\end{align}
The last step of \eqref{D19_chi} is because $z\le 1$. Inserting \eqref{D19_chi} into \eqref{D18_I21'}, we have
\begin{align}
&I_{2,1}^{\prime}\le
\frac{\left(\sigma_{2}^{2}\right)^{2}(\overline{\gamma}_{1}+\overline{\gamma}_{1} \overline{\gamma}_{2})\left(2\overline{\gamma}_{1}+2\overline{\gamma}_{1} \overline{\gamma}_{2}+\overline{\gamma}_{2}\right)}{2 \lambda_{x} \lambda_{y} \lambda_{z} \overline{\gamma}_{2} P_{\textrm{p}}^{2}}
\int_{0}^{1}
\left(2\left(\overline{\gamma}_{1}+\overline{\gamma}_{1} \overline{\gamma}_{2}\right)+\frac{\overline{\gamma}_{2}}{\sqrt{z}}\right)
e^{-z \frac{\sigma_{2}^{2}(\overline{\gamma}_{1}+\overline{\gamma}_{1}\overline{\gamma}_{2})}{\lambda_{z} \sigma_{1}^{2} \overline{\gamma}_{2}}} d z
\nonumber\\&
\le
\frac{\left(\sigma_{2}^{2}\right)^{2}(\overline{\gamma}_{1}+\overline{\gamma}_{1} \overline{\gamma}_{2})\left(2\overline{\gamma}_{1}+2\overline{\gamma}_{1} \overline{\gamma}_{2}+\overline{\gamma}_{2}\right)}{2 \lambda_{x} \lambda_{y} \lambda_{z} \overline{\gamma}_{2} P_{\textrm{p}}^{2}}
\left(\int_{0}^{1} 2\left(\overline{\gamma}_{1}+\overline{\gamma}_{1} \overline{\gamma}_{2}\right) e^{-z \frac{\sigma_{2}^{2}(\overline{\gamma}_{1}+\overline{\gamma}_{1}\overline{\gamma}_{2})}{\lambda_{z} \sigma_{1}^{2} \overline{\gamma}_{2}}} d z+\int_{0}^{1} \frac{\overline{\gamma}_{2}}{\sqrt{z}} d z\right) \triangleq I_{2,1}^{\prime \prime}.
\label{D20_I21'2}
\end{align}
Next, by using \eqref{D1_gamma1} and \eqref{D2_gamma2}, we have
\begin{align}
I_{2,1}^{\prime \prime}&=
\left(\lambda_{z} \sigma_{1}^{2} \sigma_{2}^{2}\left(1-e^{-\frac{\sigma_{2}^{2}(\overline{\gamma}_{1}+\overline{\gamma}_{1}\overline{\gamma}_{2})}{\lambda_{z} \sigma_{1}^{2} \overline{\gamma}_{2}}}\right)+\left(\sigma_{2}^{2}\right)^{2}\right)
\frac{(\overline{\gamma}_{1}+\overline{\gamma}_{1} \overline{\gamma}_{2})\left(2\overline{\gamma}_{1}+2\overline{\gamma}_{1} \overline{\gamma}_{2}+\overline{\gamma}_{2}\right)}{\lambda_{x} \lambda_{y} \lambda_{z} P_{\textrm{p}}^{2}}
\nonumber\\&
\rightarrow
\frac{2\left(\lambda_{z} \sigma_{1}^{2} \sigma_{2}^{2}+\left(\sigma_{2}^{2}\right)^{2}\right)}{\lambda_{x} \lambda_{y} \lambda_{z}}\left(\frac{\lambda_{x}}{\sigma_{1}^{2}}\right)^{2 r_{1}}\left(\frac{\lambda_{y}}{\sigma_{2}^{2}}\right)^{2 r_{2}} \frac{1}{P_{\textrm{p}}^{2-2 r_{1}-2 r_{2}}}.
\label{D21_I21''}
\end{align}
Now we consider $I_{2,2}$ in \eqref{B15_I2}. By applying the change of variables $x=\varphi_{8}(z)-\frac{x^{\prime}}{z}$ and noting that the upper limits of variables $x$ as well as $y$ approach to zero as $P_{\textrm{p}} \rightarrow \infty$, we have
\begin{align}
&I_{2,2}\rightarrow
\frac{1}{2 \lambda_{x} \lambda_{y} P_{\textrm{p}}^{2}}
\int_{0}^{\frac{\sigma_{2}^{2}(\overline{\gamma}_{1}+\overline{\gamma}_{1}\overline{\gamma}_{2})}{\sigma_{1}^{2} \overline{\gamma}_{2}}}
\frac{\left( \sqrt{\sigma_{1}^{2}\sigma_{2}^{2}\overline{\gamma}_{2}(\overline{\gamma}_{1}+\overline{\gamma}_{1} \overline{\gamma}_{2}) z}\left( 2+\sqrt{\frac{\sigma_{1}^{2}(\overline{\gamma}_{1}+\overline{\gamma}_{1} \overline{\gamma}_{2}) z}{\sigma_{2}^{2} \overline{\gamma}_{2}}}\right) +\sigma_{2}^{2} \overline{\gamma}_{2}\right)^{2}}{\left( 1+\sqrt{\frac{\sigma_{1}^{2}\left(\overline{\gamma}_{1}+\overline{\gamma}_{1} \overline{\gamma}_{2}\right) z}{\sigma_{2}^{2} \overline{\gamma}_{2}}}\right) ^{2}}
\frac{e^{-\frac{z}{\lambda_{z}}-\frac{\varphi_{8}(z)}{\lambda_{x}}}}{z \lambda_{z}} d z
\nonumber\\&\le
\frac{1}{2 \lambda_{x} \lambda_{y} P_{\textrm{p}}^{2}}
\int_{0}^{\frac{\sigma_{2}^{2}(\overline{\gamma}_{1}+\overline{\gamma}_{1}\overline{\gamma}_{2})}{\sigma_{1}^{2} \overline{\gamma}_{2}}}
\frac{\left( \sqrt{\sigma_{1}^{2}\sigma_{2}^{2}\overline{\gamma}_{2}(\overline{\gamma}_{1}+\overline{\gamma}_{1} \overline{\gamma}_{2}) z}\left( 2+2\sqrt{\frac{\sigma_{1}^{2}(\overline{\gamma}_{1}+\overline{\gamma}_{1} \overline{\gamma}_{2}) z}{\sigma_{2}^{2} \overline{\gamma}_{2}}}\right) +\sigma_{2}^{2} \overline{\gamma}_{2}\right)^{2}}{\left( 1+\sqrt{\frac{\sigma_{1}^{2}\left(\overline{\gamma}_{1}+\overline{\gamma}_{1} \overline{\gamma}_{2}\right) z}{\sigma_{2}^{2} \overline{\gamma}_{2}}}\right) ^{2}}
\frac{e^{-\frac{\sigma_{2}^{2} \overline{\gamma}_{2}}{\lambda_{x} P_{\textrm{p}} z}}}{z \lambda_{z}} d z \triangleq I_{2,2}^{\prime}.
\label{D22_I22}
\end{align}
Next, by making use of $z \leq \frac{\sigma_{2}^{2}(\overline{\gamma}_{1}+\overline{\gamma}_{1}\overline{\gamma}_{2})}{\sigma_{1}^{2} \overline{\gamma}_{2}}$, one can show that
\begin{align}
I_{2,2}^{\prime}&=
\frac{1}{2 \lambda_{x} \lambda_{y} P_{\textrm{p}}^{2}}
\int_{0}^{\frac{\sigma_{2}^{2}(\overline{\gamma}_{1}+\overline{\gamma}_{1}\overline{\gamma}_{2})}{\sigma_{1}^{2} \overline{\gamma}_{2}}}
\left( 2\sqrt{\sigma_{1}^{2}\sigma_{2}^{2}\overline{\gamma}_{2}(\overline{\gamma}_{1}+\overline{\gamma}_{1} \overline{\gamma}_{2}) z} +\frac{\sigma_{2}^{2} \overline{\gamma}_{2}}{{ 1+\sqrt{\frac{\sigma_{1}^{2}\left(\overline{\gamma}_{1}+\overline{\gamma}_{1} \overline{\gamma}_{2}\right) z}{\sigma_{2}^{2} \overline{\gamma}_{2}}}}}\right)^{2}
\frac{e^{-\frac{\sigma_{2}^{2} \overline{\gamma}_{2}}{\lambda_{x} P_{\textrm{p}} z}}}{z \lambda_{z}} d z 
\nonumber\\&\le
\frac{1}{2 \lambda_{x} \lambda_{y} P_{\textrm{p}}^{2}}
\int_{0}^{\frac{\sigma_{2}^{2}(\overline{\gamma}_{1}+\overline{\gamma}_{1}\overline{\gamma}_{2})}{\sigma_{1}^{2} \overline{\gamma}_{2}}}
\left(2 \sigma_{2}^{2}(\overline{\gamma}_{1}+\overline{\gamma}_{1} \overline{\gamma}_{2})+\sigma_{2}^{2} \overline{\gamma}_{2}\right)^{2}
\frac{e^{-\frac{\sigma_{2}^{2} \overline{\gamma}_{2}}{\lambda_{x} P_{\textrm{p}} z}}}{z \lambda_{z}} d z \triangleq I_{2,2}^{\prime \prime}.
\label{D23_I22'}
\end{align}
Furthermore, applying the change of variables $z=\frac{\sigma_{2}^{2} \overline{\gamma}_{2}}{\lambda_{x} P_{\textrm{p}} z^{\prime}}$ and utilizing \cite[Eq. (6.5.15)]{HandbookOf} as well as \cite[Eq. (5.1.11)]{HandbookOf}, one can show that 
\begin{align}
I_{2,2}^{\prime \prime}&=
\left(-C+\ln \left(\frac{\lambda_{x}\left(\overline{\gamma}_{1}+\overline{\gamma}_{1} \overline{\gamma}_{2}\right) P_{p}}{\sigma_{1}^{2}\left(\overline{\gamma}_{2}\right)^{2}}\right)-\frac{\sum_{n=1}^{\infty}\left(-\frac{\sigma_{1}^{2}\left(\overline{\gamma}_{2}\right)^{2}}{\lambda_{x}\left(\overline{\gamma}_{1}+\overline{\gamma}_{1} \overline{\gamma}_{2}\right) P_{\textrm{p}}}\right)^{n}}{n \times n !}\right)
\frac{\left(\sigma_{2}^{2}\right)^{2}\left(2 \overline{\gamma}_{1}+2 \overline{\gamma}_{1} \overline{\gamma}_{2}+\overline{\gamma}_{2}\right)^{2}}{2 \lambda_{x} \lambda_{y} \lambda_{z} P_{\textrm{p}}^{2}}
\nonumber\\&\rightarrow
\ln \left(\left(\frac{\lambda_{x}}{\sigma_{1}^{2}}\right)^{r_{1}+1}\left(\frac{\lambda_{y}}{\sigma_{2}^{2}}\right)^{-r_{2}} P_{\textrm{p}}^{1+r_{1}-r_{2}}\right) \frac{2\left(\sigma_{2}^{2}\right)^{2}}{\lambda_{x} \lambda_{y} \lambda_{z}}\left(\frac{\lambda_{x}}{\sigma_{1}^{2}}\right)^{2 r_{1}}\left(\frac{\lambda_{y}}{\sigma_{2}^{2}}\right)^{2 r_{2}} \frac{1}{P_{\textrm{p}}^{2-2 r_{1}-2 r_{2}}}
,
\label{D24_I22''}
\end{align}
where $C\approx0.5772156649$ denotes the Euler's constant. The last step in \eqref{D24_I22''} is obtained by using \eqref{D1_gamma1} and \eqref{D2_gamma2}. Now we turn to $I_{3,1}$. As $P_{\textrm{p}} \rightarrow \infty$, it follows from \eqref{B18_I3} that 
\begin{align}
I_{3,1}&\rightarrow
\int_{\frac{\sigma_{2}^{2}(\overline{\gamma}_{1}+\overline{\gamma}_{1}\overline{\gamma}_{2})}{\sigma_{1}^{2} \overline{\gamma}_{2}}}^{\infty}
\int_{0}^{\frac{\sigma_{1}^{2}\left( \overline{\gamma}_{1}+\overline{\gamma}_{1}\overline{\gamma}_{2}+\overline{\gamma}_{2}\right) }{P_{\textrm{p}}}}
\int_{0}^{x \frac{\sigma_{2}^{2}}{\sigma_{1}^{2}}} \frac{1}{\lambda_{y}} d y \frac{1}{\lambda_{x}} d x \frac{e^{-\frac{z}{\lambda_{z}}}}{\lambda_{z}} d z
=\frac{\sigma_{1}^{2} \sigma_{2}^{2}\left(\overline{\gamma}_{1}+\overline{\gamma}_{1} \overline{\gamma}_{2}+\overline{\gamma}_{2}\right)^{2} e^{-\frac{\sigma_{2}^{2}\left(\overline{\gamma}_{1}+\overline{\gamma}_{1}\overline{\gamma}_{2}\right)}{\lambda_{z} \sigma_{1}^{2} \overline{\gamma}_{2}}}}{2 \lambda_{x} \lambda_{y} P_{\textrm{p}}^{2}}
\nonumber\\&\le
\frac{\sigma_{1}^{2} \sigma_{2}^{2}\left(\overline{\gamma}_{1}+\overline{\gamma}_{1} \overline{\gamma}_{2}+\overline{\gamma}_{2}\right)^{2} }{2 \lambda_{x} \lambda_{y} P_{\textrm{p}}^{2}}
\rightarrow \frac{1}{2}\left(\frac{\lambda_{x}}{\sigma_{1}^{2}}\right)^{2 r_{1}-1}\left(\frac{\lambda_{y}}{\sigma_{2}^{2}}\right)^{2 r_{2}-1} \frac{1}{P_{\textrm{p}}^{2-2 r_{1}-2 r_{2}}}.
\label{D25_I31}
\end{align}
Finally, we consider $I_{3,2}$. By applying the change of variables $x=x^{\prime} \frac{\sigma_{2}^{2}(\overline{\gamma}_{1}+\overline{\gamma}_{1}\overline{\gamma}_{2}+\overline{\gamma}_{2})}{P_{\textrm{p}} z}+\frac{\sigma_{1}^{2}(\overline{\gamma}_{1}+\overline{\gamma}_{1}\overline{\gamma}_{2}+\overline{\gamma}_{2})}{P_{\textrm{p}} }$ and noting that the upper limits of variables $x$ and $y$ approach to zero as $P_{\textrm{p}} \rightarrow \infty$, it follows from \eqref{B18_I3} that 
\begin{align}
I_{3,2}&\rightarrow
\frac{\left(\sigma_{2}^{2}\right)^{2}\left(\overline{\gamma}_{1}+\overline{\gamma}_{1} \overline{\gamma}_{2}+\overline{\gamma}_{2}\right)^{2}}{2 \lambda_{x} \lambda_{y} \lambda_{z} P_{\textrm{p}}^{2}} \int_{\frac{\sigma_{2}^{2}\left(\overline{\gamma}_{1}+\overline{\gamma}_{1} \overline{\gamma}_{2}\right)}{\sigma_{1}^{2} \overline{\gamma}_{2}}}^{\infty} \frac{e^{-\frac{z}{\lambda_{z}}}}{z} d z
\nonumber\\&=\frac{\left(\sigma_{2}^{2}\right)^{2}\left(\overline{\gamma}_{1}+\overline{\gamma}_{1} \overline{\gamma}_{2}+\overline{\gamma}_{2}\right)^{2}}{2 \lambda_{x} \lambda_{y} \lambda_{z} P_{\textrm{p}}^{2}}
\Gamma\left(0, \frac{\sigma_{2}^{2}\left(\overline{\gamma}_{1}+\overline{\gamma}_{1} \overline{\gamma}_{2}\right)}{\lambda_{z} \sigma_{1}^{2} \overline{\gamma}_{2}}\right) \triangleq I_{3,2}^{\prime}.
\label{D26_I32}
\end{align}
By noting that $\Gamma(0,\infty)=0$ and then using \eqref{D1_gamma1} as well as \eqref{D2_gamma2}, we have
\begin{align}
I_{3,2}&\le
\frac{\left(\sigma_{2}^{2}\right)^{2}\left(\overline{\gamma}_{1}+\overline{\gamma}_{1} \overline{\gamma}_{2}+\overline{\gamma}_{2}\right)^{2}}{2 \lambda_{x} \lambda_{y} \lambda_{z} P_{\textrm{p}}^{2}}
\rightarrow \frac{\left(\sigma_{2}^{2}\right)^{2}}{2 \lambda_{x} \lambda_{y} \lambda_{z}}\left(\frac{\lambda_{x}}{\sigma_{1}^{2}}\right)^{2 r_{1}}\left(\frac{\lambda_{y}}{\sigma_{2}^{2}}\right)^{2 r_{2}} \frac{1}{P_{\textrm{p}}^{2-2 r_{1}-2 r_{2}}}.
\label{D27_I32'}
\end{align}
Note that the decaying rate of $P_{\textrm{out,BC}}^{\prime}$ is determined by the dominating term. According to the foregoing results, the decaying rates of $I_{1,2}$, $I_{2,1}$, $I_{2,2}$, $I_{3,1}$, and $I_{3,2}$ are faster than or at least the same as that of $I_{1,1}$, which decays proportionally to $\frac{1}{P_{\textrm{p}}^{2-2 r_{1}-2 r_{2}}}$. Therefore, we can conclude that the DMT performance of the BC-NOMA scheme is $(2-2r_{\textrm{A}}-2r_{\textrm{B}})$.
\bibliographystyle{IEEEtran}
\bibliography{BCNOMA}
\end{document}